\pdfoutput=1

\documentclass[11pt,twoside,a4paper,cmspaper,final,collab]{cms-tdr}
\def\svnVersion{438f7ee-D}\def\svnDate{2020/02/24}\def\cmsCernNoTag{CERN-EP-2019-124}\def\cmsCernDate{\today}\def\cmsMessage{"Published in the Journal of High Energy Physics as \href{http://dx.doi.org/10.1007/JHEP03(2020)056}{\doi{10.1007/JHEP03(2020)056}.}"}

\begin{document}\cmsNoteHeader{TOP-18-009}

\hyphenation{had-ron-i-za-tion}
\hyphenation{cal-or-i-me-ter}
\hyphenation{de-vices}

\newlength\cmsFigWidth
\ifthenelse{\boolean{cms@external}}{\setlength\cmsFigWidth{0.49\textwidth}}{\setlength\cmsFigWidth{0.65\textwidth}}
\ifthenelse{\boolean{cms@external}}{\providecommand{\cmsLeft}{upper\xspace}}{\providecommand{\cmsLeft}{left\xspace}}
\ifthenelse{\boolean{cms@external}}{\providecommand{\cmsRight}{lower\xspace}}{\providecommand{\cmsRight}{right\xspace}}

\providecommand{\cmsTable}[1]{\resizebox{\textwidth}{!}{#1}}
\newlength\cmsTabSkip\setlength{\cmsTabSkip}{1ex}
\providecommand{\PV}{\ensuremath{\mathrm{V}}}
\newcommand{\pp}{\ensuremath{\Pp\Pp}\xspace}
\newcommand{\Njets}{\ensuremath{N_\text{j}}\xspace}
\newcommand{\Nbjets}{\ensuremath{N_{\cPqb}}\xspace}
\newcommand{\mll}{\ensuremath{m(\ell\ell)}\xspace}
\newcommand{\mZ}{\ensuremath{m(\cPZ)}\xspace}
\newcommand{\mt}{\ensuremath{m(\cPqt)}\xspace}
\newcommand{\WZ}{\ensuremath{\PW\cPZ}\xspace}
\newcommand{\ZZ}{\ensuremath{\PZ\cPZ}\xspace}
\newcommand{\VVV}{\ensuremath{\PV\PV\PV}\xspace}
\newcommand{\ttZ}{\ensuremath{\ttbar\cPZ}\xspace}
\newcommand{\ttVV}{\ensuremath{\ttbar\PV\PV}\xspace}
\newcommand{\ttX}{\ensuremath{\cPqt(\cPaqt)\mathrm{X}}\xspace}
\newcommand{\ttW}{\ensuremath{\ttbar\PW}\xspace}
\newcommand{\ttH}{\ensuremath{\ttbar\PH}\xspace}
\newcommand{\tZq}{\ensuremath{\cPqt\PZ\Pq}\xspace}
\newcommand{\Xgamma}{\ensuremath{\mathrm{X}\gamma}\xspace}
\newcommand{\Zgamma}{\ensuremath{\PZ\Pgg^{(*)}}\xspace}
\newcommand{\ttgamma}{\ensuremath{\ttbar\Pgg^{(*)}}\xspace}
\newcommand{\ttgammaReal}{\ensuremath{\ttbar\Pgg}\xspace}
\newcommand{\tWZ}{\ensuremath{\cPqt\PW\PZ}\xspace}
\newcommand{\tHq}{\ensuremath{\cPqt\PH\Pq}\xspace}
\newcommand{\tHW}{\ensuremath{\cPqt\PH\PW}\xspace}

\newcommand{\nbtags}{\ensuremath{N_{\cPqb}}\xspace}
\newcommand{\njets}{\ensuremath{N_{\mathrm{j}}}\xspace}
\newcommand{\nleptons}{\ensuremath{N_{\ell}}\xspace}

\newcommand{\pTZ}{\ensuremath{\pt(\PZ)}\xspace}
\newcommand{\cosThetaStar}{\ensuremath{\cos\theta^\ast_{\PZ}}\xspace}

\newcommand{\coupling}[2]{\ensuremath{C_{#1,\cmsSymbolFace{#2}}}}
\newcommand{\ConeV}{\coupling{1}{V}}
\newcommand{\ConeA}{\coupling{1}{A}}
\newcommand{\CtwoV}{\coupling{2}{V}}
\newcommand{\CtwoA}{\coupling{2}{A}}

\newcommand{\cpt}{\ensuremath{c_{\phi\PQt}}\xspace}
\newcommand{\cpQM}{\ensuremath{c_{\phi \mathrm{Q}}^{-}}\xspace}
\newcommand{\ctZ}{\ensuremath{c_{\PQt\PZ}}\xspace}
\newcommand{\ctZI}{\ensuremath{c_{\PQt\PZ}^{[I]}}\xspace}

\newcommand{\muR}{\ensuremath{\mu_R}\xspace}
\newcommand{\muF}{\ensuremath{\mu_F}\xspace}

\newcommand{\correlated}{\ensuremath{\checkmark}\xspace}
\newcommand{\uncorrelated}{\ensuremath{\times}\xspace}

\newcommand{\thetaw}{\ensuremath{\theta_{\PW}}}
\newcommand{\sinw}{\ensuremath{\sin\thetaw}}
\newcommand{\cosw}{\ensuremath{\cos\thetaw}}
\newcommand{\swcw}{\ensuremath{\sinw\cosw}}

\cmsNoteHeader{TOP-18-009}

\title{Measurement of top quark pair production in association with a \PZ boson in proton-proton collisions at $\sqrt{s}=13\TeV$}

\date{\today}

\abstract{A measurement of the inclusive cross section of top quark pair production in association with a \PZ boson using proton-proton collisions at a center-of-mass energy of 13\TeV at the LHC is performed.  The data sample corresponds to an integrated luminosity of 77.5\fbinv, collected by the CMS experiment during 2016 and 2017. The measurement is performed using final states containing three or four charged leptons (electrons or muons), and the \PZ boson is detected through its decay to an oppositely charged lepton pair. The production cross section is measured to be  $\sigma(\ttZ)=0.95\pm0.05\stat\pm0.06\syst\unit{pb}$. For the first time, differential cross sections are measured as functions of the transverse momentum of the \PZ boson and the angular distribution of the negatively charged lepton from the \PZ boson decay. The most stringent direct limits to date on the anomalous couplings of the top quark to the \PZ boson are presented, including constraints on the Wilson coefficients in the framework of the standard model effective field theory. }

\hypersetup{%
pdfauthor={CMS Collaboration},%
pdftitle={Measurement of top quark pair production in association with a Z boson in proton-proton collisions at sqrt(s)=13 TeV},%
pdfsubject={CMS},%
pdfkeywords={CMS, physics, top quark, Z boson}}

\maketitle

\section{Introduction}
\label{sec:Introduction}

The large amount of proton-proton (\pp) collision data at a center-of-mass energy of 13\TeV at the CERN LHC allows for precision measurements of standard model (SM) processes with very small production rates.
Precise measurements of the inclusive and differential cross sections of the \ttZ process are of particular interest because it can receive sizable contributions from phenomena beyond the SM (BSM)~\cite{Bylund2016,Englert2016}.  The \ttZ production is the most sensitive process for directly measuring the coupling of the top quark to the \PZ boson. Also, this process is an important background to several searches for BSM phenomena, as well as to measurements of certain SM processes, such as \ttbar production in association with the Higgs boson (\ttH).

 {\tolerance=500
The inclusive cross section for \ttZ production has been measured by both the CMS and ATLAS Collaborations using \pp collision data at $\sqrt{s}=13\TeV$, corresponding to an integrated luminosity of about 36\fbinv.
The CMS Collaboration used events containing three or four charged leptons~(muons or electrons) collected in 2016 and reported a value $\sigma(\ttZ)=0.99^{+0.09}_{-0.08}\stat\,^{+0.12}_{-0.10}\syst\unit{pb}$~\cite{Sirunyan:2017uzs}. The ATLAS Collaboration used events with two, three, or four charged leptons in a data sample collected in 2015 and 2016 and measured $\sigma(\ttZ)=0.95\pm 0.08\stat\pm 0.10\syst\unit{pb}$~\cite{Aaboud:2019njj}.
\par}

In this paper, we report an updated measurement of the \ttZ cross section in three- and four-lepton final states using \pp collision data collected with the CMS detector in 2016 and 2017, corresponding to a total integrated luminosity of 77.5\fbinv.
The \PZ boson is detected through its decay to an oppositely charged lepton pair.
While the data analysis strategy remains similar to the one presented in Ref.~\cite{Sirunyan:2017uzs}, this new measurement benefits largely from an improved lepton selection procedure based on multivariate analysis techniques and a more inclusive trigger selection.
In addition to the inclusive cross section, the differential cross section is measured as a function of the  transverse momentum of the \PZ boson, \pTZ,  and \cosThetaStar. The latter observable is the cosine of the angle between the direction of the \PZ boson in the detector reference frame and the direction of the negatively charged lepton in the rest frame of the \PZ boson.

Because of the key role of the top quark interaction with the \PZ boson in many BSM models~\cite{Hollik:1998vz,Agashe:2006wa,Kagan:2009bn,Ibrahim:2010hv,Ibrahim:2011im,Grojean:2013qca}, the differential cross section measurements can be used to constrain anomalous \ttZ couplings.
To this end, we pursue two different interpretations. A Lagrangian containing anomalous couplings~\cite{AguilarSaavedra:2008zc} is used to obtain bounds on the vector and axial-vector currents, as well as on the electroweak magnetic and electric dipole moments of the top quark. The interpretation is extended in the context of SM effective field theory (SMEFT)~\cite{Grzadkowski:2010es}, and we constrain the Wilson coefficients of the relevant BSM operators of mass dimension~6. There are 59 operators, among which we select the four most relevant linear combinations, as described in Ref.~\cite{AguilarSaavedra:2018nen}.

This paper is organized as follows.
In Section~\ref{sec:cms}, a brief description of the CMS detector is provided.
In Section \ref{sec:objects}, the simulation of signal and background processes is discussed, followed by the description of the selection of events online (during data taking) and
offline (after data taking) in Section~\ref{sec:eventselection}.
The background estimation is discussed in Section~\ref{sec:backgrounds}, and
the sources of systematic uncertainties that affect the measurements are discussed in Section~\ref{sec:Systematic}.
In Section~\ref{sec:Results}, we present the results of the inclusive and differential measurements, followed by the limits on anomalous couplings and SMEFT interpretation.
The results are summarized in Section~\ref{sec:Conclusions}.

\section{The CMS detector}
\label{sec:cms}

The central feature of the CMS apparatus is a superconducting solenoid of 6\unit{m} internal diameter, providing a magnetic field of 3.8\unit{T}. Within the solenoid volume are a silicon pixel and strip tracker, a lead tungstate crystal electromagnetic calorimeter (ECAL), and a brass and scintillator hadron calorimeter, each composed of a barrel and two endcap sections. Forward calorimeters extend the pseudorapidity ($\eta$) coverage.
Muons are detected in gas-ionization chambers embedded in the steel magnetic flux-return yoke outside the solenoid.  Events of interest are selected using a two-tiered trigger system~\cite{Khachatryan:2016bia}. The first level, composed of custom hardware processors, uses information from the calorimeters and muon detectors to select events, while the second level
selects events by running a version of the full event reconstruction software optimized for fast processing on a farm of computer processors. A more detailed description of the CMS detector, together with a definition of the coordinate system used and the relevant kinematic variables, can be found in Ref.~\cite{Chatrchyan:2008zzk}.

\section {Data samples and object selection}
\label{sec:objects}

The data sample used in this measurement corresponds to an integrated luminosity of 77.5\fbinv of \pp collision events collected with the CMS detector during 2016 and 2017.
To incorporate the LHC running conditions and the CMS detector performance, the two data sets were analyzed independently with appropriate calibrations applied, and combined at the final stage to extract the cross section value, as described in more detail in Section~\ref{sec:Systematic}.

Simulated Monte Carlo (MC) events are used to model the signal selection efficiency, to test the background prediction techniques, and to predict some of the background yields.
Two sets of simulated events for each process are used in order to match the different data-taking conditions in 2016 and 2017.
Events for the \ttZ signal process and a variety of background processes, including production of \WZ and triple vector boson (\VVV) events, are simulated at next-to-leading order (NLO) in  perturbative quantum chromodynamics (QCD) using the \MGvATNLO~v2.3.3 and v2.4.2 generators~\cite{Alwall:2014hca}. In these simulations, up to one additional jet is included in the matrix element calculation. The NLO \POWHEG~v2~\cite{powheg2} generator is used for simulation of the \ttbar production process, as well as for processes involving the Higgs boson produced in vector boson fusion (VBF) or in association with vector bosons or top quarks.
The NNPDF3.0 (NNPDF3.1) \cite{Ball:2014uwa,Ball:2017nwa} parton distribution functions (PDFs) are used for simulating the hard process.
Table~\ref{table:samples} gives an overview of the event generators, PDF sets, and cross section calculations that are used for the signal and background processes.
For all processes, the parton showering and hadronization are simulated using \PYTHIA 8.203~\cite{Sjostrand:2007gs,Sjostrand:2014zea}.
The modeling of the underlying event is done using the CUETP8M1~\cite{Skands:2014pea,CMS-PAS-GEN-14-001} and CP5 tunes~\cite{Sirunyan:2019dfx} for simulated samples corresponding to the 2016 and 2017 data sets, respectively.
The CUETP8M2 and CUETP8M2T4 tunes~\cite{CMS-PAS-TOP-16-021} are used for the 2016 \ttH and \ttVV samples, respectively.
Double counting of the partons generated with \MGvATNLO and \PYTHIA is removed using the \textsc{FxFx}~\cite{Frederix:2012ps} matching schemes for NLO samples.

The \ttZ cross section measurement is performed in a phase space defined by
the invariant mass of an oppositely charged and same-flavor lepton pair $70\le\mll\le110\GeV$.
Using a simulated signal sample, the contribution of $\ttbar\Pgg^{*}$ was verified to be negligible.
The \PZ boson branching fractions to charged and neutral lepton pairs are set to $(\Z\to\ell\ell,\nu\nu)=0.301$~\cite{Tanabashi:2018oca}.
The theoretical prediction of the inclusive \ttZ cross section is computed for $\sqrt{s}=13\TeV$ at NLO in QCD and electroweak accuracy using \MGvATNLO and the PDF4LHC recommendations~\cite{Butterworth:2015oua} to assess the uncertainties. It is found to be $0.84\pm 0.10$\unit{pb}~\cite{deFlorian:2016spz,Frixione:2015zaa,Frederix:2018nkq}, with the renormalization and factorization scales \muF and \muR set to $\muR=\muF=\mt+\mZ/2$, where $\mt=172.5\GeV$ is the on-shell top quark mass~\cite{deFlorian:2016spz}.

\begin{table}[htb]
\centering
\topcaption{
Event generators used to simulate events for the various processes.
For each of the simulated processes shown, the order of the cross section normalization, the event generator used, the perturbative order of the generator calculation, and the NNPDF versions at NLO and at next-to-next-to-leading order (NNLO) used in simulating samples for the 2016 (2017) data sets.
}
\label{table:samples}
\cmsTable{
\begin{tabular}{ccccc}
\multirow{2}{*}{Process} & Cross section & \multirow{2}{*}{Event generator} & Perturbative & \multirow{2}{*}{NNPDF version} \\
 & normalization & & order & \\ \hline \noalign{\vskip\cmsTabSkip}
\ttZ, \tZq, \ttW, \WZ, $\PZ$+jets, & \multirow{2}{*}{NLO} & \MGvATNLO & \multirow{2}{*}{NLO} & \multirow{2}{*}{3.0 NLO (3.1 NNLO)} \\
\VVV, \ttgamma, $\PW\Pgg^{(*)}$, \Zgamma & & v2.2.3 (v2.4.2) & & \\[\cmsTabSkip]
\multirow{2}{*}{$\Pg \Pg \to \ZZ$} & \multirow{2}{*}{NLO \cite{Caola:2015psa}} & \MCFM v7.0.1 \cite{Campbell:2010ff} & \multirow{2}{*}{LO} & \multirow{2}{*}{3.0 LO (3.1LO)} \\
 & & {\small\textsc{JHUGen}}~v7.0.11~\cite{Bolognesi:2012mm} &\\[\cmsTabSkip]
$\qqbar \to \ZZ$ & NNLO \cite{Cascioli:2014yka}& \POWHEG~v2 \cite{Melia:2011tj,Nason:2013ydw} & NLO & 3.0 NLO (3.1 NNLO) \\[\cmsTabSkip]
\multirow{2}{*}{$\PW \PH$, $\PZ \PH$} & \multirow{2}{*}{NLO} & \POWHEG~v2 \textsc{minlo HVJ}~\cite{Luisoni:2013kna} & \multirow{2}{*}{NLO} & \multirow{2}{*}{3.0 NLO (3.1 NNLO)} \\
 & & {\small\textsc{JHUGen}}~v7.0.11~\cite{Bolognesi:2012mm} &\\[\cmsTabSkip]
VBF \PH & NLO & \POWHEG~v2 & NLO & 3.0 NLO (3.1 NNLO) \\[\cmsTabSkip]
\ttH & NLO & \POWHEG~v2 \cite{Hartanto:2015uka} & NLO & 3.0 NLO (3.1 NNLO) \\[\cmsTabSkip]
\ttbar & NNLO+NNLL \cite{Czakon:2011xx} & \POWHEG~v2 & NLO & 3.0 NLO (3.1 NNLO) \\[\cmsTabSkip]
\ttVV, \tHW, \tHq, \tWZ & LO & \MGvATNLO & LO & 3.0 LO (3.1 NNLO)
\end{tabular}
}
\end{table}

All events are processed through a simulation of the CMS detector based on \GEANTfour~\cite{Geant} and are reconstructed with the same algorithms as used for data.
Minimum-bias \pp interactions occuring in the same or nearby bunch crossing, referred to as pileup (PU), are also simulated, and the observed distribution of the reconstructed \pp interaction vertices in an event is used to ensure that the simulation describes the data.
The CMS particle-flow (PF) algorithm~\cite{Sirunyan:2017ulk} is used for particle reconstruction and identification, yielding a consistent set of electron~\cite{Khachatryan:2015hwa}, muon~\cite{Chatrchyan:2012xi}, charged and neutral hadron, and photon candidates.
These particles are defined with respect to the primary IV (PV), chosen to have the largest value of summed physics-object $\pt^2$, where these physics objects are reconstructed by a jet-finding algorithm~\cite{Cacciari:2008gp,Cacciari:2011ma} applied to all charged tracks associated with the vertex.
Jets are reconstructed by clustering PF candidates using the anti-\kt algorithm~\cite{Cacciari:2008gp} with a distance parameter $R=0.4$.
The influence of PU is mitigated through a charged hadron subtraction technique, which removes the energy of charged hadrons not originating from the PV~\cite{CMS-PAS-JME-14-001}.
Jets are calibrated separately in simulation and data, accounting for energy deposits of neutral particles from PU and any nonlinear detector response~\cite{Chatrchyan:2011ds,Khachatryan:2016kdb}.
Jets with $\pt> 30\GeV$ and $\abs{\eta}<2.4$ are selected for the analysis.
Jets are identified as originating from the hadronization of \cPqb quarks using the \textsc{DeepCSV} algorithm~\cite{Sirunyan:2017ezt}.
This algorithm achieves an averaged efficiency of 70\% for \cPqb quark jets to be correctly identified, with a misidentification rate of 12\% for charm quark jets and 1\% for jets originating from \cPqu, \cPqd, \cPqs quarks or gluons.

Lepton identification and selection are critical ingredients in this measurement.
Prompt leptons are those originating from direct \PW or \PZ boson decays, while nonprompt are those that are either misidentified jets or genuine leptons resulting from semileptonic decays of hadrons containing heavy-flavor quarks.
To achieve an effective rejection of the nonprompt leptons, a multivariate analysis has been developed separately for electrons and muons similar to the one presented in Ref.~\cite{Sirunyan:2018hoz}.
A boosted decision tree (BDT) classifier is used via the TMVA toolkit~\cite{Hocker:2007ht} for the multivariate analysis.
In addition to the lepton \pt and $\abs{\eta}$, the training uses several discriminating variables.
These comprise the kinematic properties of the jet closest to the lepton; the impact parameter in the transverse plane of the lepton track with respect to the PV; a variable that quantifies the quality of the geometric matching of the track in the silicon tracker with the signals measured in the muon chambers; variables related to the ECAL shower shape of electrons; two variants of relative isolation---one computed with a fixed ($R=0.3$) and another with a variable cone size depending on the lepton \pt~\cite{Khachatryan:2016uwr}.
The relative isolation is defined as the scalar \pt sum of the particles within a cone around the lepton direction, divided by the lepton \pt.
Comparing a stringent requirement on the BDT output to the non-BDT-based lepton identification used in Ref.~\cite{Sirunyan:2017uzs}, an increase of up to 15\% in prompt lepton selection efficiency is achieved, while the nonprompt lepton selection efficiency is reduced by about a factor 2 to 4, depending on the lepton \pt.
Muons~(electrons) passing the BDT selection and having  $\pt> 10\GeV$ and $\abs{\eta}<2.4~(2.5)$ are selected.
The efficiency for prompt leptons in the \ttZ signal events in the three lepton channel is around 90\% when averaged over \pt range used in the analysis for both electrons and muons. In the four-lepton channel, a less stringent lepton selection is used and it results in an average efficiency of 95\%.
In order to avoid double counting, jets within a cone of $\Delta R=\sqrt{\smash[b]{(\Delta\eta)^2+(\Delta\phi)^2}}=0.4$ around the selected leptons are discarded, where $\Delta\eta$ and $\Delta\phi$ are the differences in pseudorapidity and azimuthal angle, respectively.

\section{Event selection and observables}
\label{sec:eventselection}

Events are selected using a suite of triggers each of which requires the presence of one, two, or three leptons.
For events selected by the triggers that require at least one muon or electron, the \pt threshold for muons (electrons) was 24 (27)\GeV during 2016 and 27 (32)\GeV in 2017.
For triggers that require the presence of at least two leptons, the \pt thresholds are 23 and 17\GeV for the highest \pt (leading) and 12 and 8\GeV for the second-highest \pt (subleading) electron and muon, respectively.
This strategy ensures an overall trigger efficiency higher than 98\% for events passing the lepton selection described below over the entire 2016 and 2017 data sets.
These efficiencies are measured in data samples with an independent trigger selection and compared to those obtained in simulation.
The measured differences are mitigated by reweighting the simulation by appropriate factors that differ from unity by less than 2~(3)\% in the 2016~(2017) data set.

Events with exactly three leptons ($\Pgm\Pgm\Pgm$, $\Pgm\Pgm\Pe$, $\Pgm\Pe\Pe$, or $\Pe\Pe\Pe$) satisfying $\pt >40, 20, 10\GeV$ or exactly four leptons ($\Pgm\Pgm\Pgm\Pgm$, $\Pgm\Pgm\Pgm\Pe$, $\Pgm\Pgm\Pe\Pe$, $\Pgm\Pe\Pe\Pe$, or $\Pe\Pe\Pe\Pe$) with $\pt > 40$, 10, 10, 10\GeV are analyzed separately.
In both categories, exactly one oppositely charged and same-flavor lepton pair consistent with the \PZ boson hypothesis is required, namely, for the three- and four-lepton categories $\abs{\mll - \mZ} <10$ and $20\GeV$, respectively. This selection reduces the contributions from background events with zero or more than one \PZ boson. Events containing zero jets are rejected.
The measurement uses the jet multiplicity \Njets in different event categories depending on the number of \cPqb-tagged jets \Nbjets in the event. For the three-lepton channel these are $\Nbjets= 0, 1, \ge 2 $, while for the four-lepton channel these categories are limited to $\Nbjets=0, \ge 1$.
The analysis makes use of several control regions in data to validate the background predictions, as well as to control the systematic uncertainties associated with them. The details are given in Section~\ref{sec:backgrounds}.

\section{Background predictions}
\label{sec:backgrounds}

Several SM processes contribute to the three- and four-lepton final states.
The \ttZ process typically produces events with large jet and \cPqb-tagged jet multiplicities.
In contrast, events with $\Nbjets=0$ are dominated by background processes.
Following closely the methodologies used in~Ref.~\cite{Sirunyan:2017uzs}, the separation between signal and backgrounds is obtained from a binned maximum-likelihood fit with nuisance parameters. In the fit, the contributions from the various background processes are allowed to vary within their uncertainties.

The main contributions to the background arise from processes with at least one top quark produced in association with a \PW, \PZ, or Higgs boson, \ie, \ttH, \ttW, \tWZ, \tZq, \tHq, \tHW, \ttVV, and $\ttbar\ttbar$.
They are collectively denoted as \ttX and estimated using simulated samples.
We consider both the theoretical and experimental systematic uncertainties in the background yields for the \ttX category.
The theoretical uncertainty in the inclusive cross section is evaluated by varying \muR and \muF in the matrix element and parton shower description by a factor of 2 up and down, ignoring the anticorrelated variations, as well as the uncertainties stemming from the choice of PDFs.
For each of these processes, this uncertainty is found to be not larger than 11\%~\cite{Campbell:2013yla, Frixione:2015zaa, Alwall:2014hca}. Among them, the \tZq cross section was recently measured by the CMS Collaboration with a precision of $15\%$~\cite{Sirunyan:2018zgs}.
Thus, we use this measurement and its uncertainty for the \tZq cross section, and 11\% as uncertainty for the normalization of the other processes.

The \WZ production constitutes the second-largest background contribution, in particular for events with three leptons, while in the four-lepton category, \ZZ production becomes substantial. For both these processes, the prediction of the overall production rate and the relevant kinematic distributions can be validated in data samples that do not overlap with the signal region.
Events with three leptons, two of which form a same-flavor pair with opposite charge and  satisfy $\abs{\mll- \mZ} <  10\GeV$ and $\Nbjets=0$, are used to validate the \WZ background prediction.
Four-lepton events with two \PZ boson candidates are used to constrain the uncertainties in the prediction of the \ZZ yield.
\begin{figure}[h!]
\centering
\includegraphics[width=0.49\textwidth]{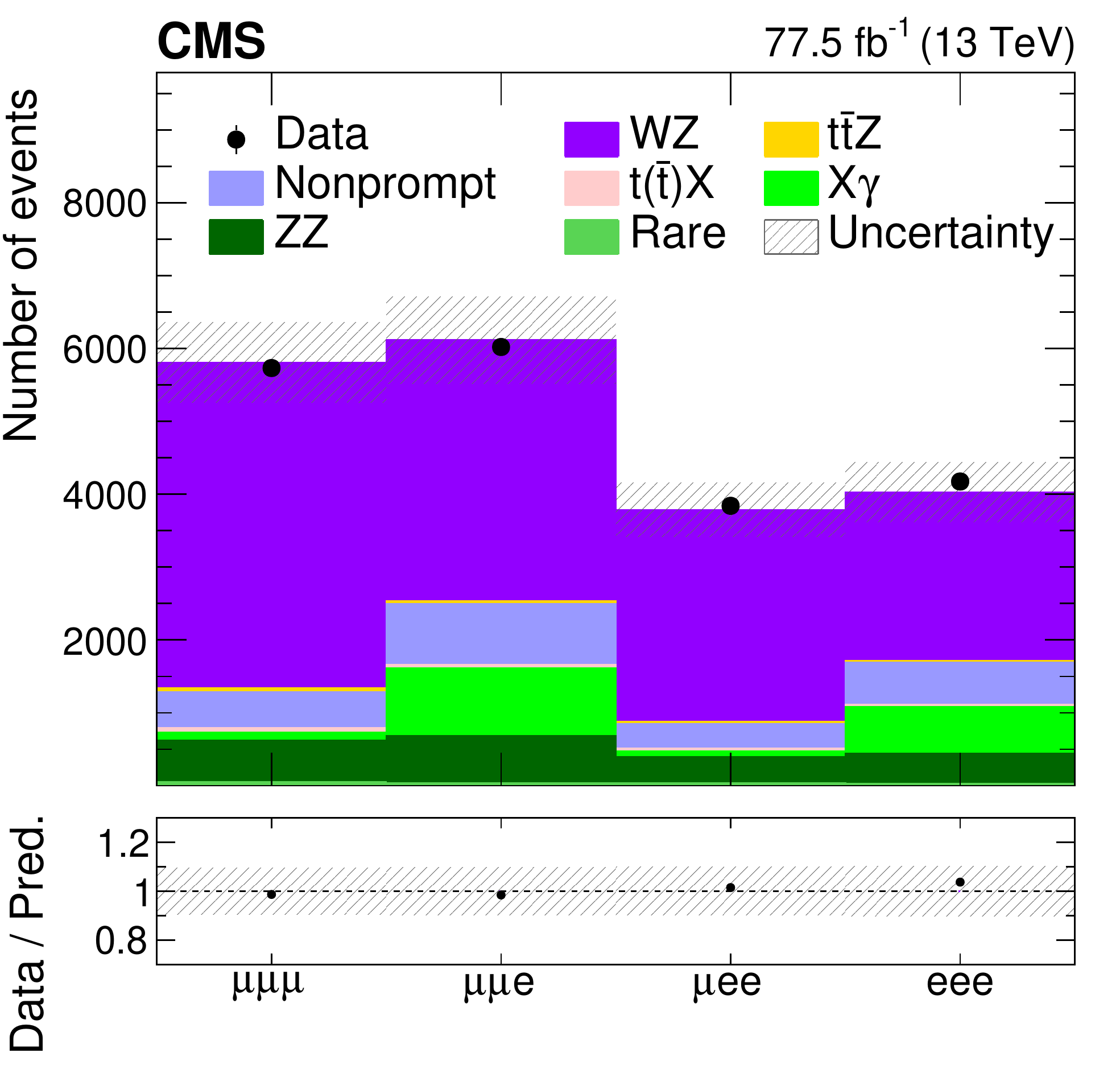}
\includegraphics[width=0.49\textwidth]{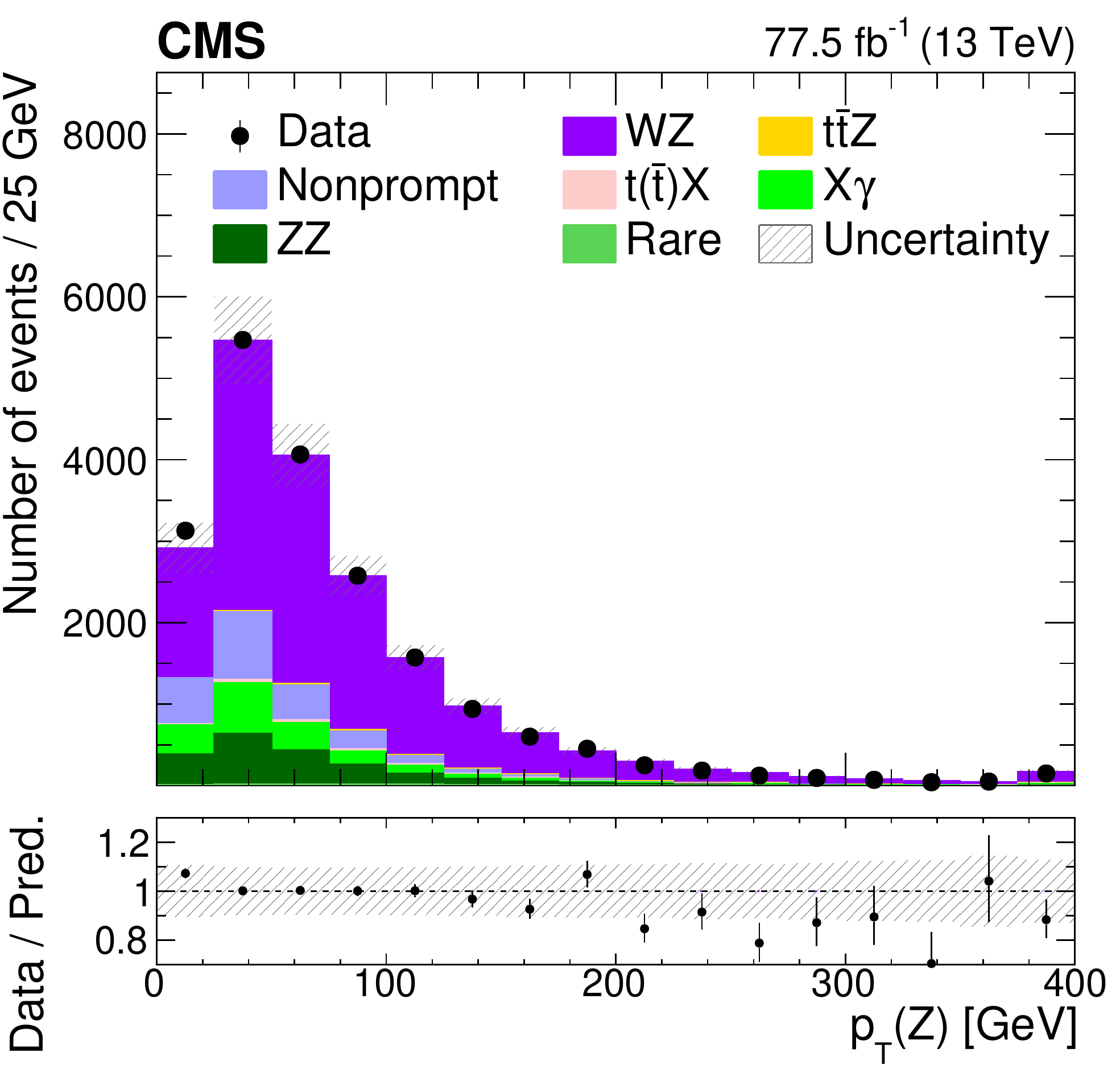}\\
\includegraphics[width=0.49\textwidth]{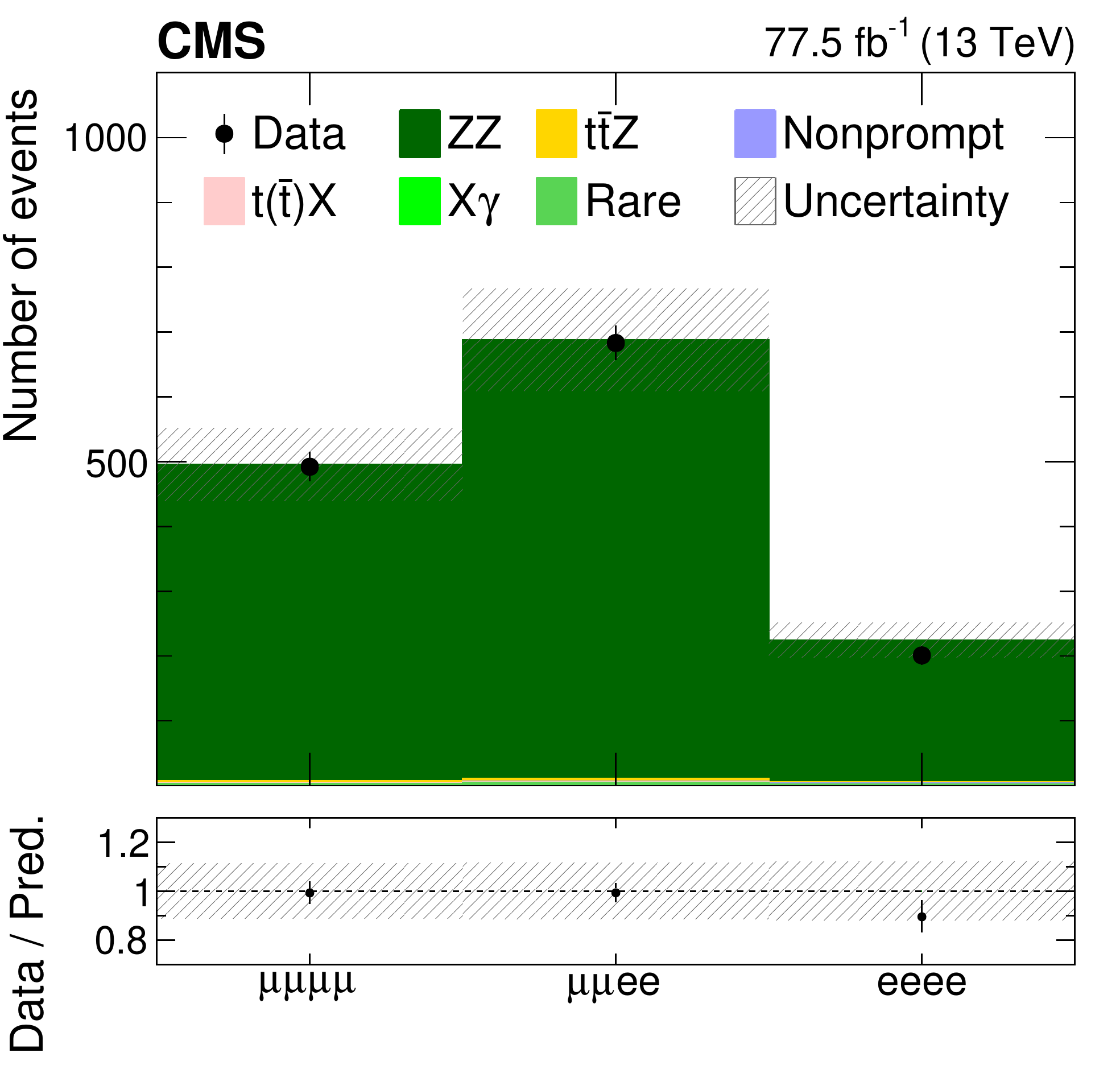}
\includegraphics[width=0.49\textwidth]{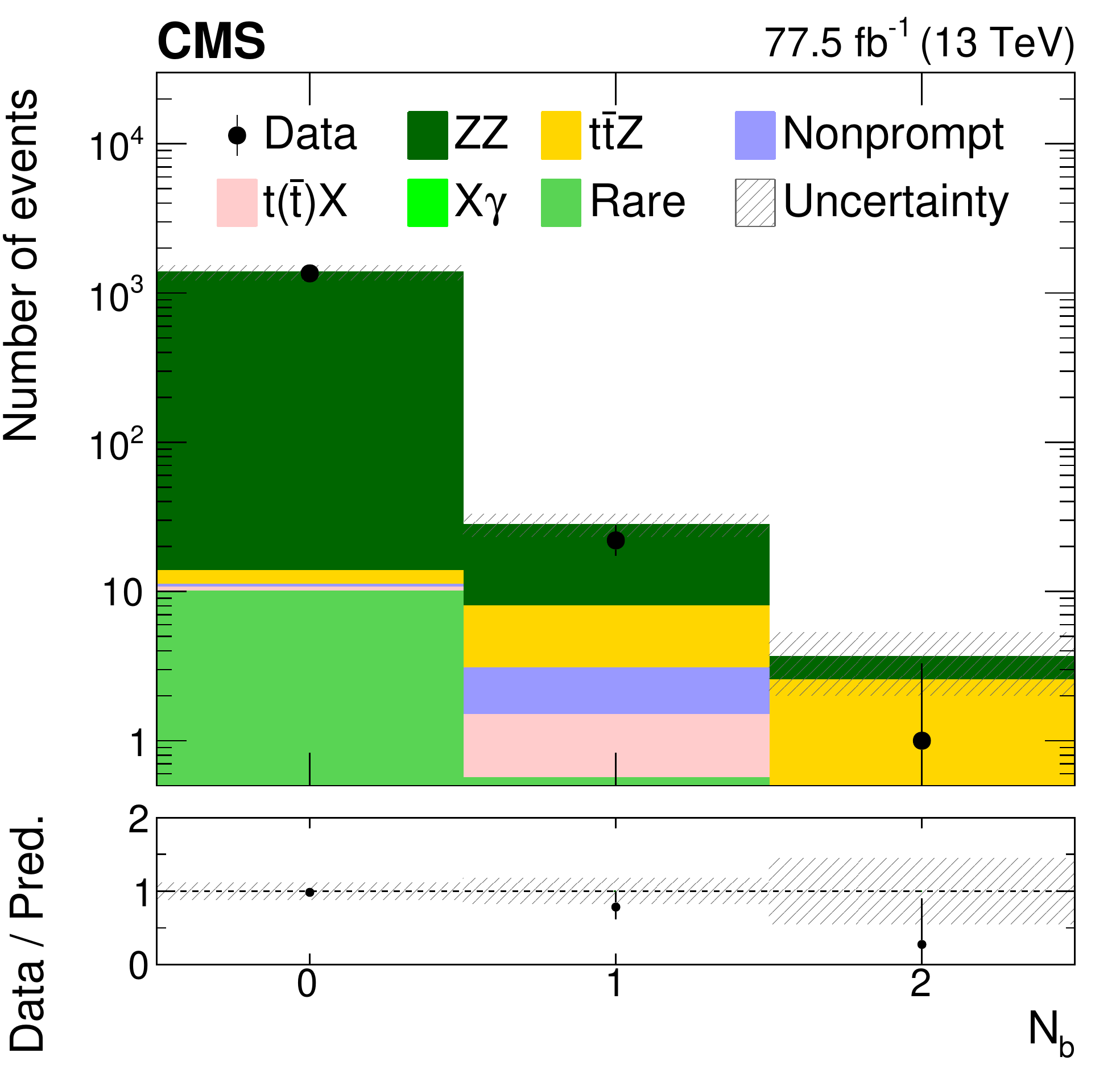}
\caption{The observed (points) and predicted (shaded histograms) event yields versus lepton flavor (upper left), and the reconstructed transverse momentum of the \PZ boson candidates (upper right) in the \WZ-enriched data control event category, and versus lepton flavor (lower left) and \Nbjets (lower right) in the \ZZ-enriched event category.
The vertical lines on the points show the statistical uncertainties in the data, and the band the total uncertainty in the predictions. The lower panels show the ratio of the event yields in data to the predictions.}
\label{figures:WZ_background}
\end{figure}

\begin{figure}[h!]
\centering
\includegraphics[width=0.49\textwidth]{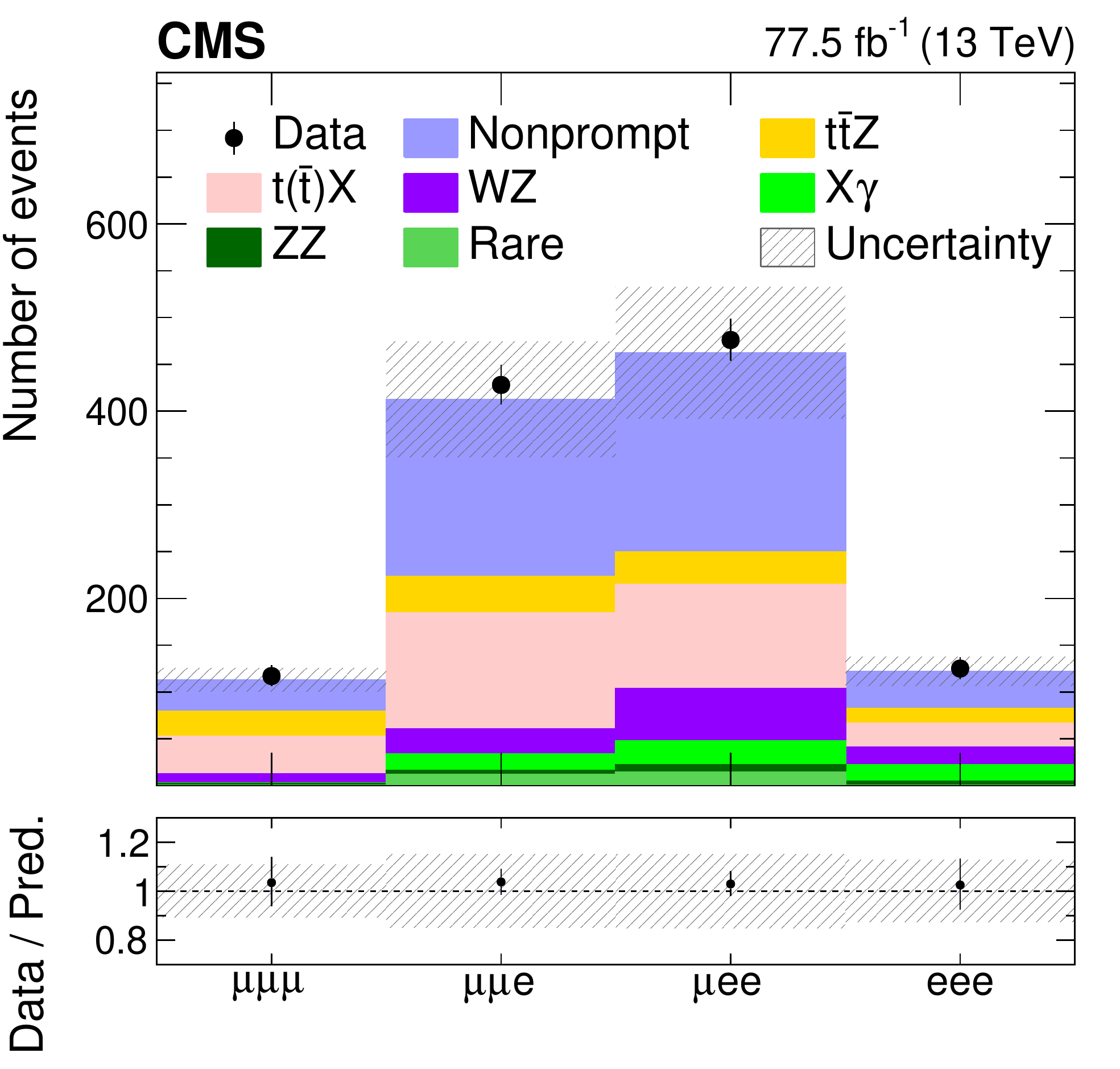}
\includegraphics[width=0.49\textwidth]{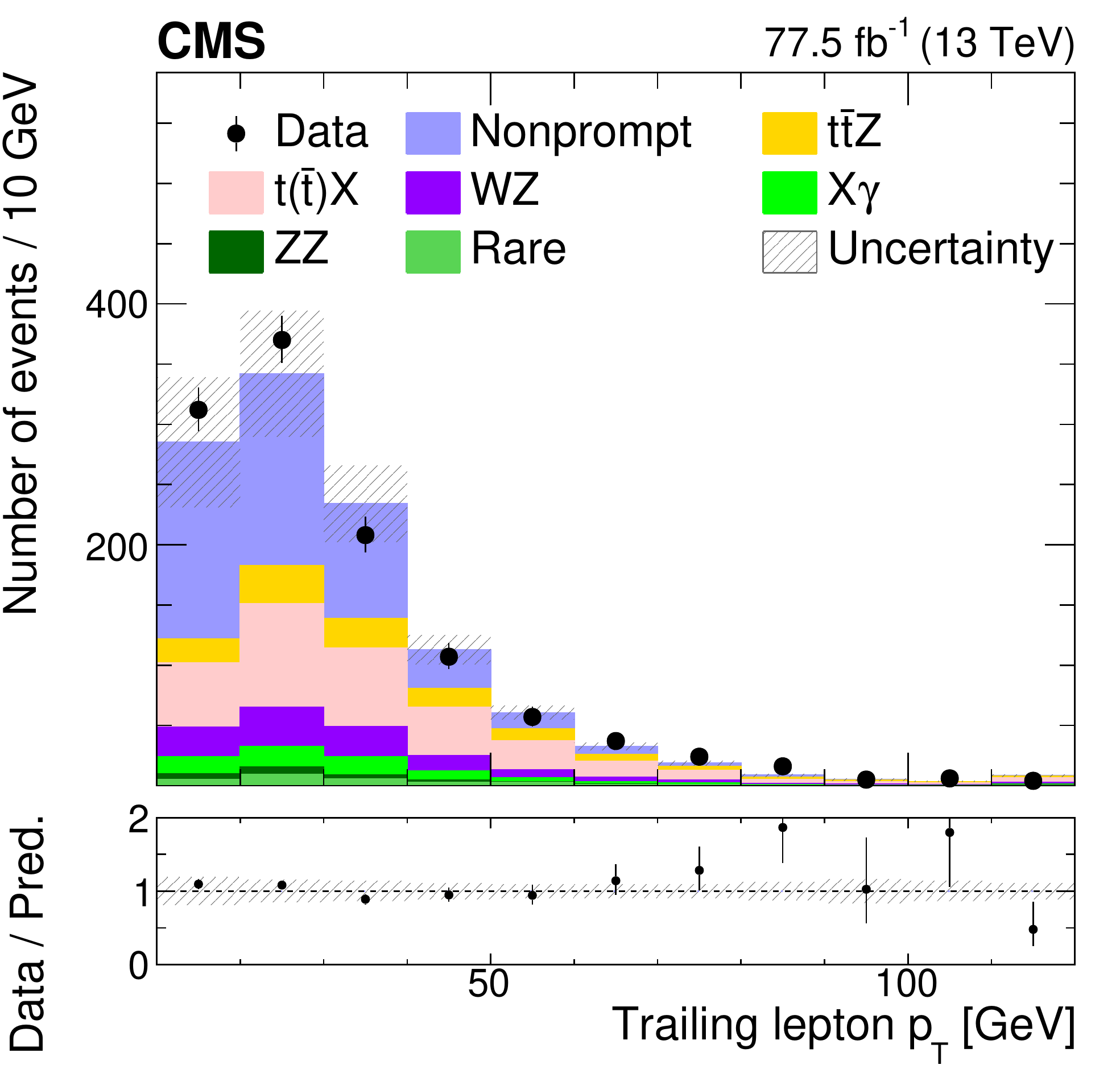}\\
\includegraphics[width=0.49\textwidth]{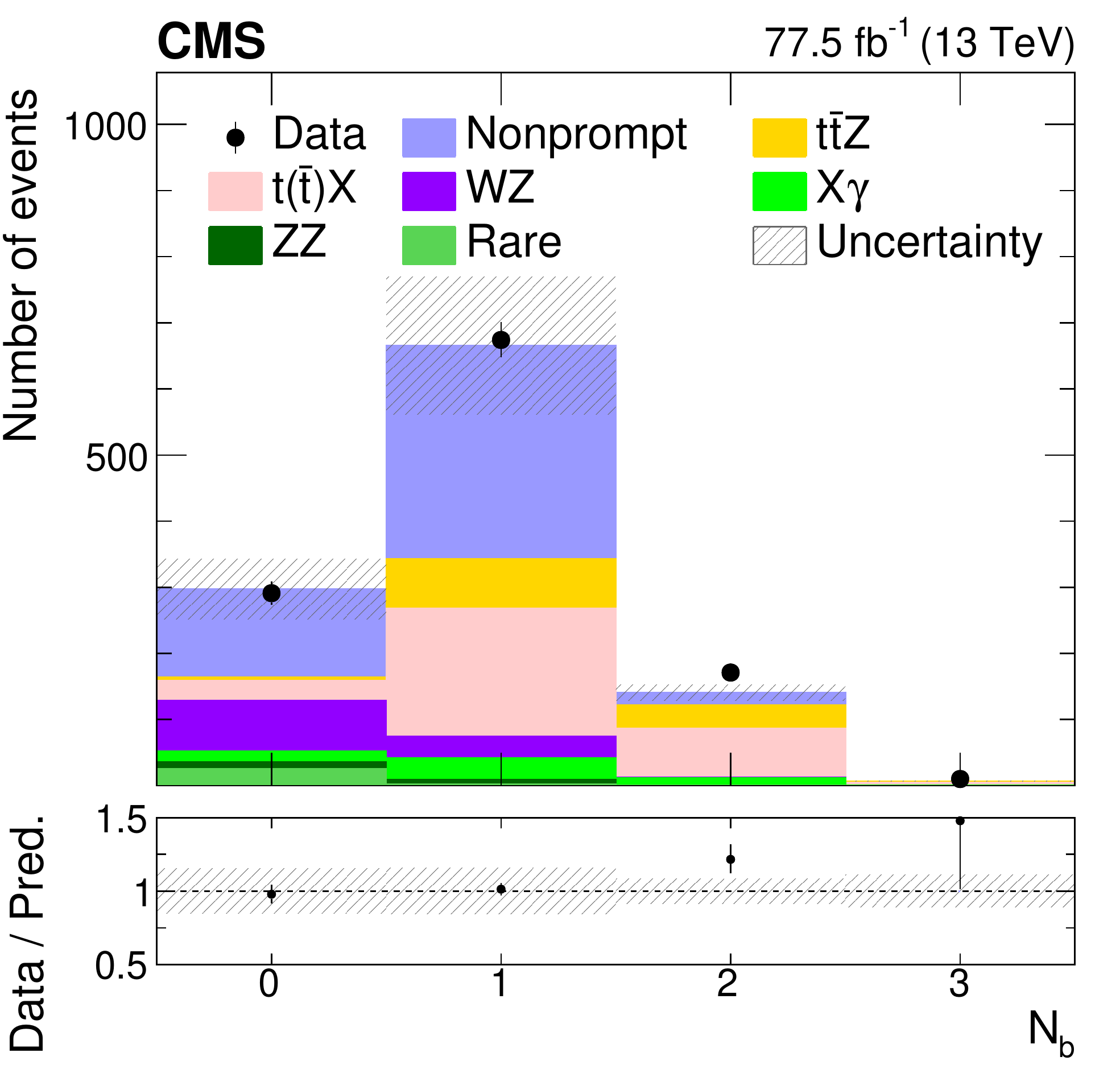}
\caption{The observed (points) and predicted (shaded histograms) event yields in regions enriched with nonprompt lepton
backgrounds in \ttbar-like processes as a function of the lepton flavors (upper left), the \pt of the lowest-\pt (trailing) lepton (upper right),
and \Nbjets (bottom).
The vertical lines on the points show the statistical uncertainties in the data, and the band the total uncertainty in the predictions. The lower panels show the ratio of the event yields in data to the background predictions. }
\label{figures:ttbar3L_background}
\end{figure}

Figure~\ref{figures:WZ_background} presents the observed and predicted event yields for these categories and the reconstructed transverse momentum of the \PZ boson candidates, as well as the lepton flavor and \Nbjets in the \ZZ-enriched control region.
Agreement within the systematic uncertainties is observed.
A normalization uncertainty of 10\% is assigned to the prediction of the \WZ and \ZZ backgrounds~\cite{Sirunyan:2018vkx,Sirunyan:2019bez}, and an additional 20\% uncertainty is appended to the \WZ background prediction with $\Njets \ge 3$ because of the observed discrepancy in events with high jet multiplicity.

We also estimate the potential mismodeling of \WZ production when heavy-quark pairs from gluon splitting are included by using a control data sample containing a \PZ boson candidate and two \cPqb-tagged jets.
The distribution of the angle between the two \cPqb jets is sensitive to the modeling of gluon splitting and good agreement is observed. A systematic uncertainty of 20\% is estimated from possible mismodeling.
Taking into account the fraction of simulated \WZ events with gluon splitting, the additional uncertainty in the prediction of \WZ events with $\Nbjets \geq 1$ is estimated to be 8\%.

The background with nonprompt leptons mainly originates from \ttbar or $\PZ \to \ell \ell$ events in which a nonprompt lepton arises from a semileptonic decay of a heavy-flavor hadron or misidentified jets in addition to two prompt leptons.
The lepton selection specifically targets the reduction of nonprompt-lepton backgrounds to a subdominant level, while keeping the signal efficiency high.
The details of the nonprompt-lepton background estimation are given in Ref.~\cite{Sirunyan:2017uzs}.
In this analysis, it is validated in simulation and with a data control sample that contains three-lepton events without a \PZ boson candidate.
Figure~\ref{figures:ttbar3L_background} shows the predicted and observed yields in this control sample for different lepton flavors, as a function of the \pt of the lowest-\pt lepton and \Nbjets.
We find good agreement between predicted and observed yields.
Based on these studies, a systematic uncertainty of 30\% in the prediction of the background with nonprompt leptons is assigned, while the statistical uncertainty ranges between 5--50\%, depending on the measurement bin.

A small contribution to the background comes from \VVV processes.
We group them in the ``rare'' category as these have relatively small production rates. Processes that involve a photon (\Zgamma and \ttgammaReal ) are denoted by \Xgamma. The contribution from both of these categories to the selected event count is evaluated using simulated samples described in Section~\ref{sec:objects}.
As in the case of the \ttX backgrounds, scale factors are applied to account for small differences between data and simulation in trigger selection, lepton identification, jet energy corrections, and \cPqb jet selection efficiency.  The overall uncertainty in the normalization  of the rare  background category is estimated to be 50\%~\cite{deFlorian:2016spz,Nhung:2013jta}, while for \Xgamma it is 20\%~\cite{Khachatryan:2015kea, Khachatryan:2017jub}. The statistical uncertainty stemming from the finite size of the simulated background samples are typically small, around 5\% and reaching 100\% only in the highest jet multiplicity regions.
The simulation of photon conversion is validated in a data sample with three-lepton events where the invariant mass of the three leptons is required to be consistent with the \PZ boson mass.
Good agreement between data and simulation is observed.

\section{Systematic uncertainties}
\label{sec:Systematic}
The systematic uncertainties affecting the signal selection efficiency and background yields are summarized in Table~\ref{table:systematics}.
The table shows the range of variations in the different bins of the analysis caused by each systematic uncertainty on the signal and background yields, as well as an estimate of the impact of each input uncertainty on the measured cross section.

The table also indicates whether the uncertainties are treated as uncorrelated or fully correlated between the 2016 and 2017 data sets.

\begin{table}[h!]
\topcaption{Summary of the sources, magnitudes, treatments, and effects of the systematic uncertainties in the final \ttZ cross section measurement. The first column indicates the source of the uncertainty, the second column shows the corresponding input uncertainty range for each background source and the signal. The third column indicates how correlations are treated between the uncertainties in the 2016 and 2017 data, where \correlated means fully correlated and \uncorrelated~uncorrelated.
The last column gives the corresponding systematic uncertainty in the \ttZ cross section using the fit result.
The total systematic uncertainty, the statistical uncertainty and the total uncertainty in the \ttZ cross section are shown in the last three lines. }
\label{table:systematics}
\centering
\begin{tabular}{lccc}
\multirow{2}{*}{Source} & Uncertainty & Correlated between & Impact on the \ttZ \\
& range (\%) & 2016 and 2017 & cross section (\%) \\ \hline
Integrated luminosity & 2.5 & \uncorrelated & 2 \\
PU modeling & 1--2 & \correlated & 1 \\
Trigger & 2 & \uncorrelated & 2 \\
Lepton ID efficiency & 4.5--6 & \correlated & 4 \\
Jet energy scale & 1--9 & \correlated & 2 \\
Jet energy resolution & 0--1 & \correlated & $<$1 \\
\cPqb tagging light flavor & 0--4 & \uncorrelated & $<$1 \\
\cPqb tagging heavy flavor & 1--4 & \uncorrelated & 2 \\
Choice in \muR and \muF & 1--4 & \correlated & 1 \\
PDF choice & 1--2 & \correlated & $<$1 \\
Color reconnection & 1.5 & \correlated & 1 \\
Parton shower & 1--8 & \correlated & $<$1 \\
\WZ cross section & 10 & \correlated & 3 \\
\WZ high jet multiplicity & 20 & \correlated & 1 \\
\WZ + heavy flavor & 8 & \correlated & 1 \\
\ZZ cross section & 10 & \correlated & 1 \\
\ttX background & 10--15 & \correlated & 2 \\
X$\gamma$ background & 20 & \correlated & 1 \\
Nonprompt background & 30 & \correlated & 1 \\
Rare SM background & 50 & \correlated & 1 \\
Stat. unc. in nonprompt bkg. & 5--50 & \uncorrelated  & $<$1\\
Stat. unc. in rare SM bkg. & 5--100 & \uncorrelated   & $<$1\\[\cmsTabSkip]
Total systematic uncertainty & & & 6\\
Statistical uncertainty & & & 5\\[\cmsTabSkip]
Total & & & 8
\end{tabular}
\end{table}

The uncertainty in the integrated luminosity measurement in the 2016 (2017) data set is 2.5 (2.3)\%~\cite{CMS-PAS-LUM-17-001,CMS-PAS-LUM-17-004}, and is uncorrelated between the two data sets.
Simulated events are reweighted according to the distribution of the number of interactions in each bunch crossing corresponding to a total inelastic \pp cross section of 69.2\unit{mb}~\cite{Sirunyan:2018nqx}.
The uncertainty in the latter, which affects the PU estimate, is 5\%~\cite{ATLAS:2016pu} and leads to about 2\% uncertainty in the expected yields.

The uncertainties in the corrections to the trigger selection efficiencies are propagated to the results.
A 2\% uncertainty is assigned to the yields obtained in simulation.
Lepton selection efficiencies are measured using a ``tag-and-probe'' method~\cite{Chatrchyan:2012xi,Khachatryan:2015hwa} in bins of lepton \pt and $\eta$, and are found to be higher than 60\,(95)\% for lepton $\pt\le25~(>25)\GeV$. These measurements are performed separately in data and simulation. The differences between these two measurements are used to scale the yields obtained in the simulation.
They are typically around 1\% and reach 10\% for leptons with $\pt<20\GeV$.
The systematic uncertainties related to this source vary between 4.5 and 6\% in the signal and background yields.

Uncertainties in the jet energy calibration are estimated by shifting the jet energy corrections in simulation up and down by one standard deviation.
Depending on $\pt$ and $\eta$, the uncertainty in jet energy scale changes by 2--5\%~\cite{Khachatryan:2016kdb}.
For the signal and backgrounds modeled via simulation, the uncertainty in the measurement is determined from the observed differences in the yields with and without the shift in jet energy corrections.
The same technique is used to calculate the uncertainties from the jet energy resolution, which are found to be less than 1\%~\cite{Khachatryan:2016kdb}.
The \cPqb tagging efficiency in the simulation is corrected using scale factors determined from data~\cite{Chatrchyan:2012jua,Sirunyan:2017ezt}.
These are estimated separately for correctly and incorrectly identified jets, and each results in an uncertainty of about 1--4\%, depending on \Nbjets.

To estimate the theoretical uncertainties from the choice of \muR and \muF, each of these parameters is varied independently up and down by a factor of 2, ignoring the case, in which one parameter is scaled up while the other is scaled down.
The envelope of the acceptance variations is taken as the systematic uncertainty in each search bin and is found to be smaller than 4\%.
The different sets in the NNPDF3.0 PDF~\cite{Ball:2014uwa} are used to estimate the corresponding uncertainty in the acceptance for the differential cross section measurement, which is typically less than 1\%.
The uncertainty associated with the choice of PDFs for the anomalous coupling and SMEFT interpretations is estimated by using several PDFs and
assigning the maximum differences as the quoted uncertainty, following the PDF4LHC prescription with the MSTW2008 68\% \CL NNLO, CT10 NNLO, and NNPDF2.3 5f FFN PDF sets (as described in Ref.~\cite{Butterworth:2015oua} and references therein, as well as Refs.~\cite{Ball:2012cx,Martin:2009bu,Gao:2013xoa}).
In the parton shower simulation, the uncertainty from the choice of \muF is estimated by varying the scale of initial- and final-state radiation up by factors of 2 and $\sqrt{2}$ and down by factors of 0.5 and $1/\sqrt{2}$, respectively, as suggested in Ref.~\cite{Skands:2014pea}.
The default configuration in \PYTHIA includes a model of color reconnection based on multiple parton interactions (MPI) with early resonance decays switched off.
To estimate the uncertainty from this choice of model, the analysis is repeated with three other color reconnection models within \PYTHIA: the MPI-based scheme with early
resonance decays switched on, a gluon-move scheme~\cite{Argyropoulos:2014zoa}, and a QCD-inspired scheme~\cite{Christiansen:2015yqa}.
The total uncertainty from color reconnection modeling is estimated by taking the maximum deviation from
the nominal result and amounts to 1.5\%.

\section {Results}
\label{sec:Results}
\subsection{Inclusive cross section measurement}

\begin{figure}[h!t]
\centering{
  \includegraphics[width=.95\textwidth]{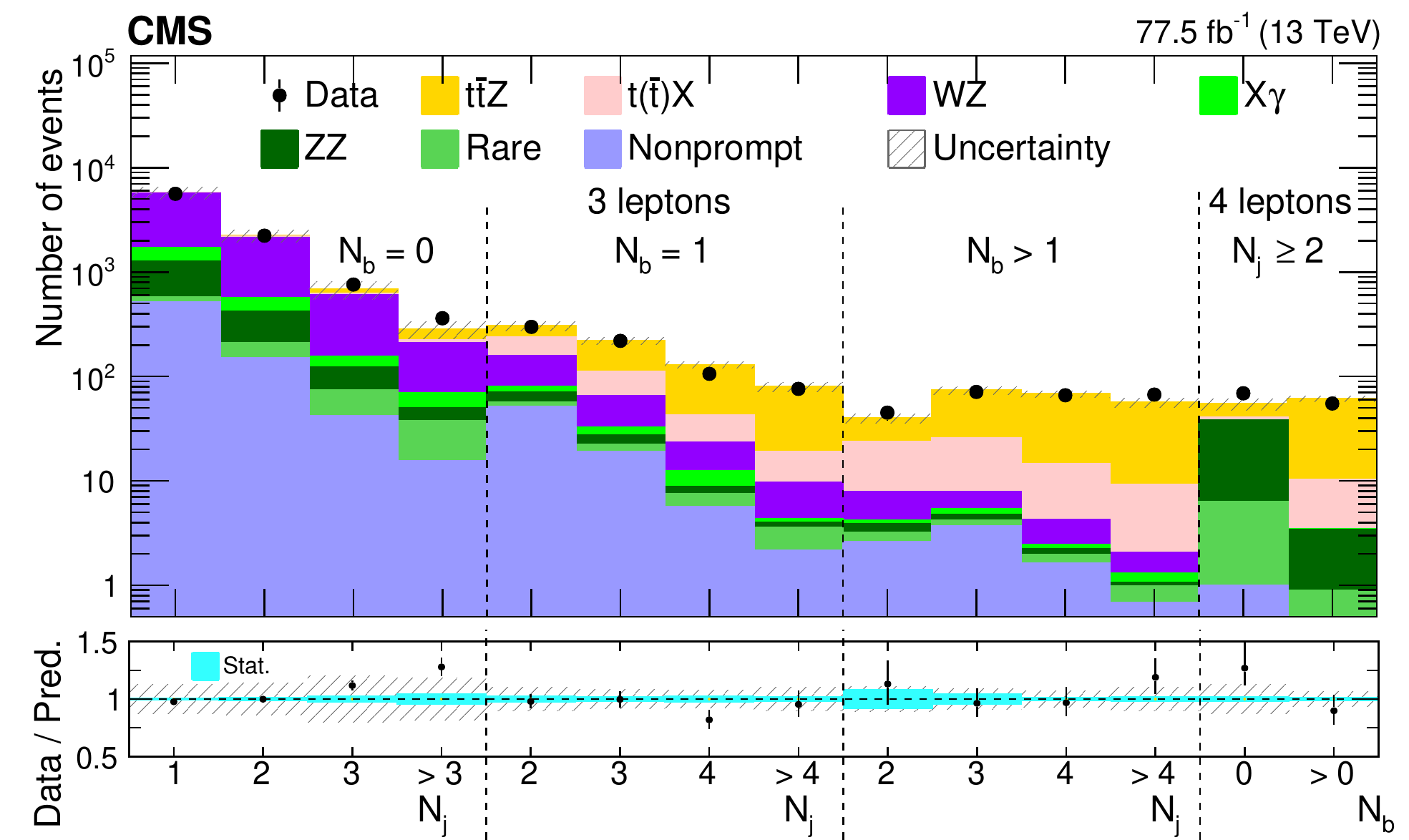}}
  \caption{Observed event yields in data for different values of \Njets and \Nbjets for events with 3 and 4 leptons, compared with the signal and background yields, as obtained from the fit. The lower panel displays the ratio of the data to the predictions of the signal and background from simulation. The inner and outer bands show the statistical and total uncertainties, respectively.
}
\label{fig:ttZ_comb}
\end{figure}

The observed data, as well as the predicted signal and background yields, are shown in Fig.~\ref{fig:ttZ_comb} in various jet and \cPqb jet categories, for events with three and four leptons.
The signal cross section is extracted from these categories using the statistical procedure detailed in Refs.~\cite{Junk:1999kv, Read:2002hq, ATL-PHYS-PUB-2011-011, Cowan:2010js}.
The observed yields and background estimates in each analysis category, and
the systematic uncertainties are used to construct a binned likelihood function $L(r, \theta)$
as a product of Poisson probabilities of all bins. As described in Section~\ref{sec:Systematic}, the bins of the two data-taking periods are kept separate, and the correlation pattern of the uncertainty as specified in Table~\ref{table:systematics}.
The parameter $r$ is the signal strength modifier, \ie, the ratio
between the measured cross section and the central value of the cross section predicted by simulation, and $\theta$ represents the full suite of nuisance parameters.

The test statistic is the profile likelihood ratio, $q(r)=-2\ln L(r,\hat\theta_{\text{r}})/L(\hat{r}, \hat{\theta})$, where $\hat\theta_{\text{r}}$ reflects the values of the nuisance parameters that maximize the likelihood function for signal strength $r$. An asymptotic approximation is used to extract the observed cross section of the signal process and the associated uncertainties~\cite{Junk:1999kv, Read:2002hq, ATL-PHYS-PUB-2011-011, Cowan:2010js}. The quantities $\hat{r}$ and $\hat{\theta}$ are the values that simultaneously maximize $L$.
The fitting procedure is performed for the inclusive cross section measurements, and separately for the SMEFT interpretation.
The combined cross section of the three- and four-lepton channels within the phase space $70\le\mll\le110$\GeV for the $\ell\ell$ pair is measured to be
\begin{equation*}
\sigma(\Pp\Pp \to \ttZ)=0.95\pm0.05\stat\pm0.06\syst\unit{pb},
\end{equation*}
in agreement with the SM prediction of $0.84\pm 0.10$\unit{pb} at NLO and electroweak accuracy~\cite{deFlorian:2016spz,Frixione:2015zaa,Frederix:2018nkq} and $0.86^{+0.07}_{-0.08}\,(\text{scale})\pm 0.03\,(\textrm{PDF}+\alpS)$\unit{pb} including also next-to-next-to-leading-logarithmic (NNLL) corrections~\cite{Kulesza:2018tqz}.
The measured cross sections for the three- and four-lepton channels are given in Table~\ref{tab:bd3L4L}.

\begin{table}[ht!]
\centering
\topcaption{The measured \ttZ cross section for events with 3 and 4 leptons and the combined measurement. \label{tab:bd3L4L}}
\begin{tabular}{cc}
Lepton requirement & Measured cross section \\[\cmsTabSkip] \hline \noalign{\vskip\cmsTabSkip}
3$\ell$            & $0.97\pm0.06\stat\pm0.06\syst\unit{pb}$ \\[\cmsTabSkip]
4$\ell$            & $0.91\pm0.14\stat\pm0.08\syst\unit{pb}$ \\[\cmsTabSkip]
Total              & $0.95\pm0.05\stat\pm0.06\syst\unit{pb}$
\end{tabular}
\end{table}

The background yields and the systematic uncertainties obtained from the fit are, in general, very close to their initial values.
The uncertainties associated with the \WZ background are modelled using three separate nuisance parameters as described in Section~\ref{sec:backgrounds}.
Events in the $\Nbjets=0$ categories provide a relatively pure \WZ control region, which helps constraining two of these uncertainties: the overall normalization uncertainty and the uncertainty in the \WZ yields with high jet multiplicity.
These uncertainties get constrained, respectively, by 30 and 70\% relative to their input values.
The third uncertainty controls the \WZ production with heavy-flavour jets populating the regions with $\Nbjets\geq 1$, and is not substantially constrained in the fit.
The individual contributions to the total systematic uncertainty in the measured cross section are listed in the fourth column of Table~\ref{table:systematics}.  The largest contribution comes from the imperfect knowledge of the lepton selection efficiencies in the signal acceptance.
The uncertainties in parton shower modeling and \ttX and \WZ background yields also form a large fraction of the total uncertainty.
With respect to the earlier measurements~\cite{Aaboud:2019njj,Sirunyan:2017uzs}, the statistical (systematic) uncertainty in the inclusive cross section is reduced by about 35~(40)\%.
The improvement in the systematic uncertainty is primarily the result of a better lepton selection procedure and the detailed studies of its performance in simulation, and an
improved estimation of the trigger and \cPqb tagging efficiencies in simulation.
The reported result is the first experimental measurement that is more precise than the most precise theoretical calculations for \ttZ production at NLO in QCD.

\begin{figure}[hp!]
  \centering
{\includegraphics[width=0.46\textwidth]{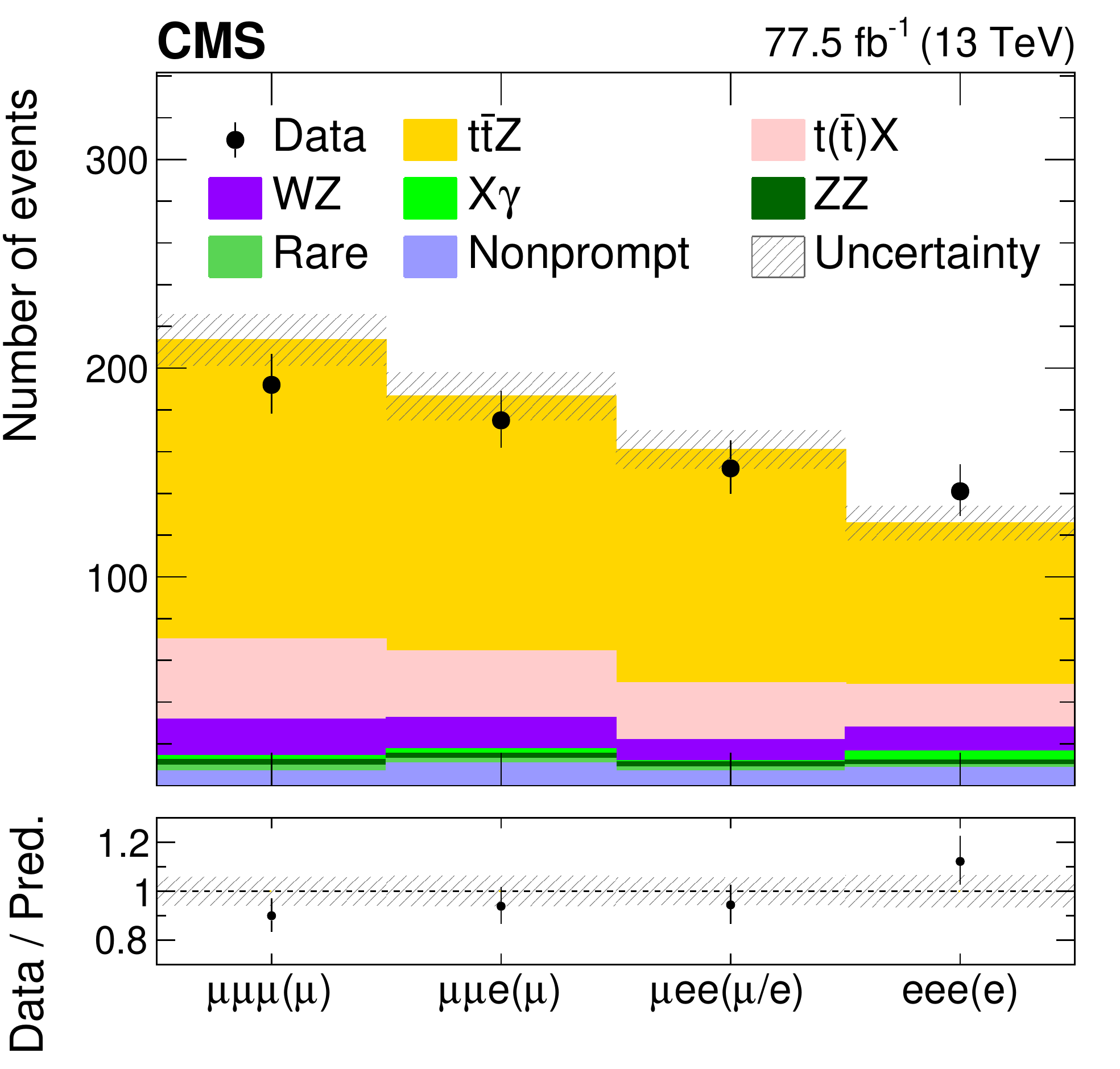}}
{\includegraphics[width=0.46\textwidth]{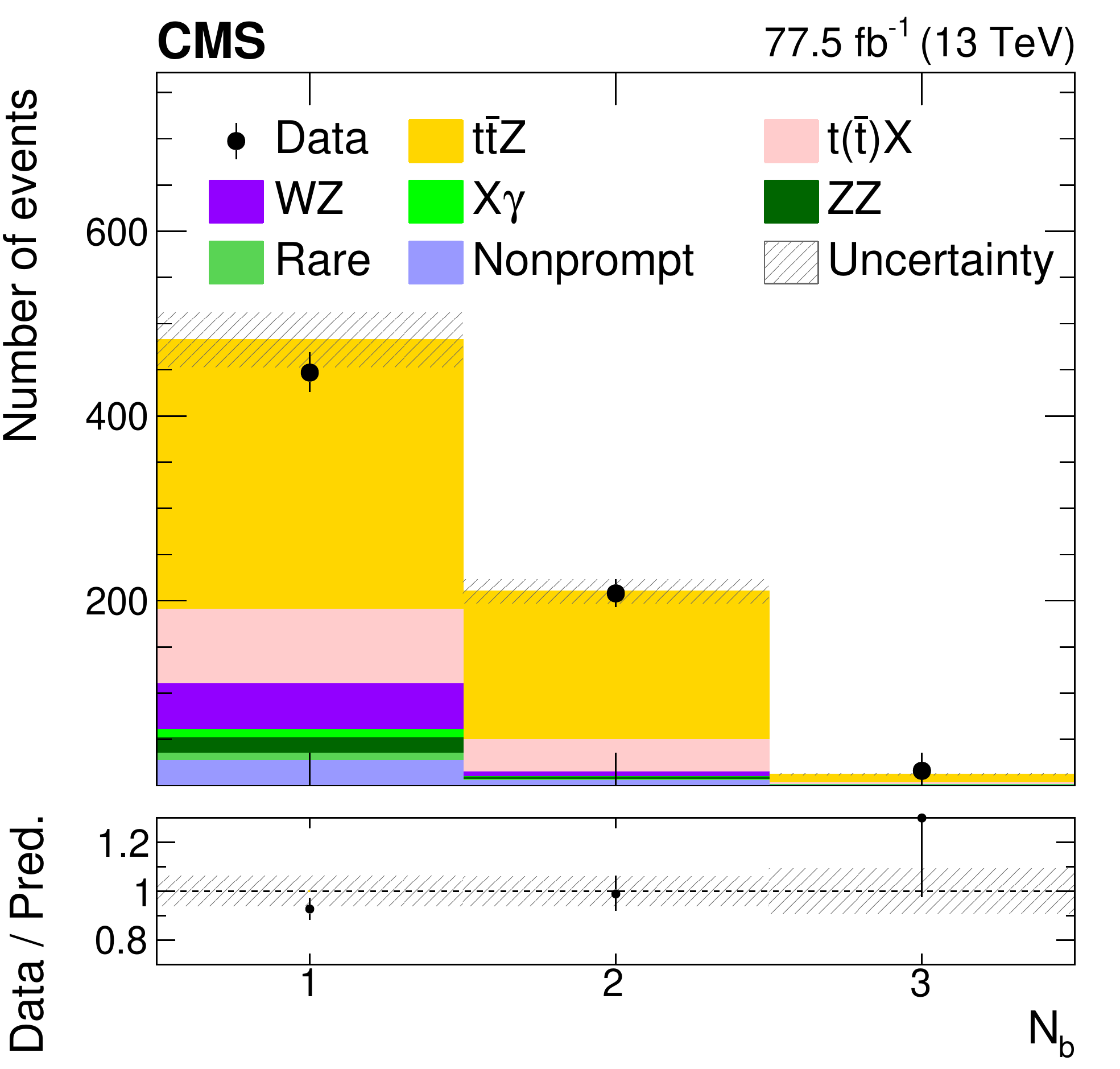}}\\
{\includegraphics[width=0.46\textwidth]{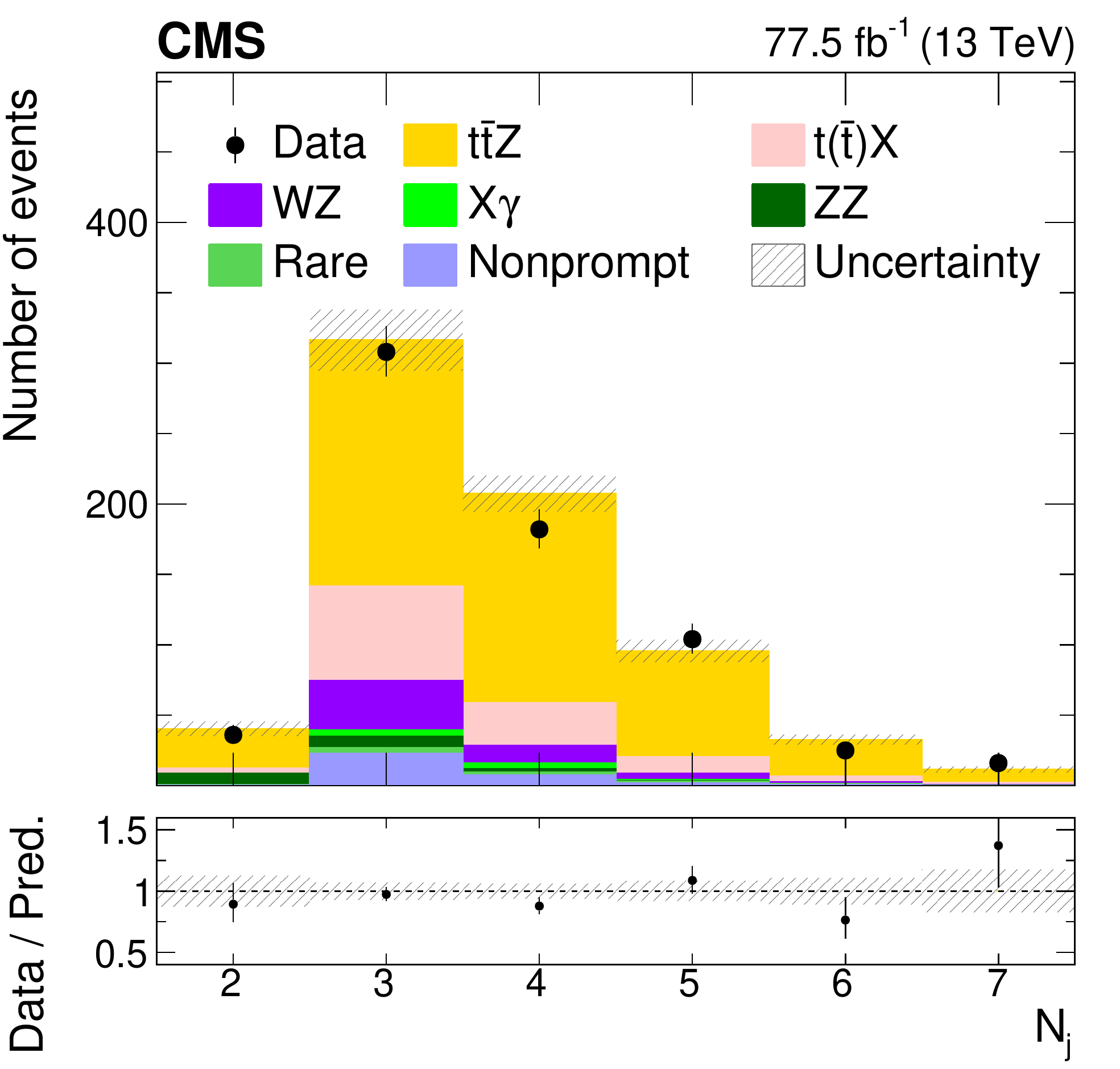}}
{\includegraphics[width=0.46\textwidth]{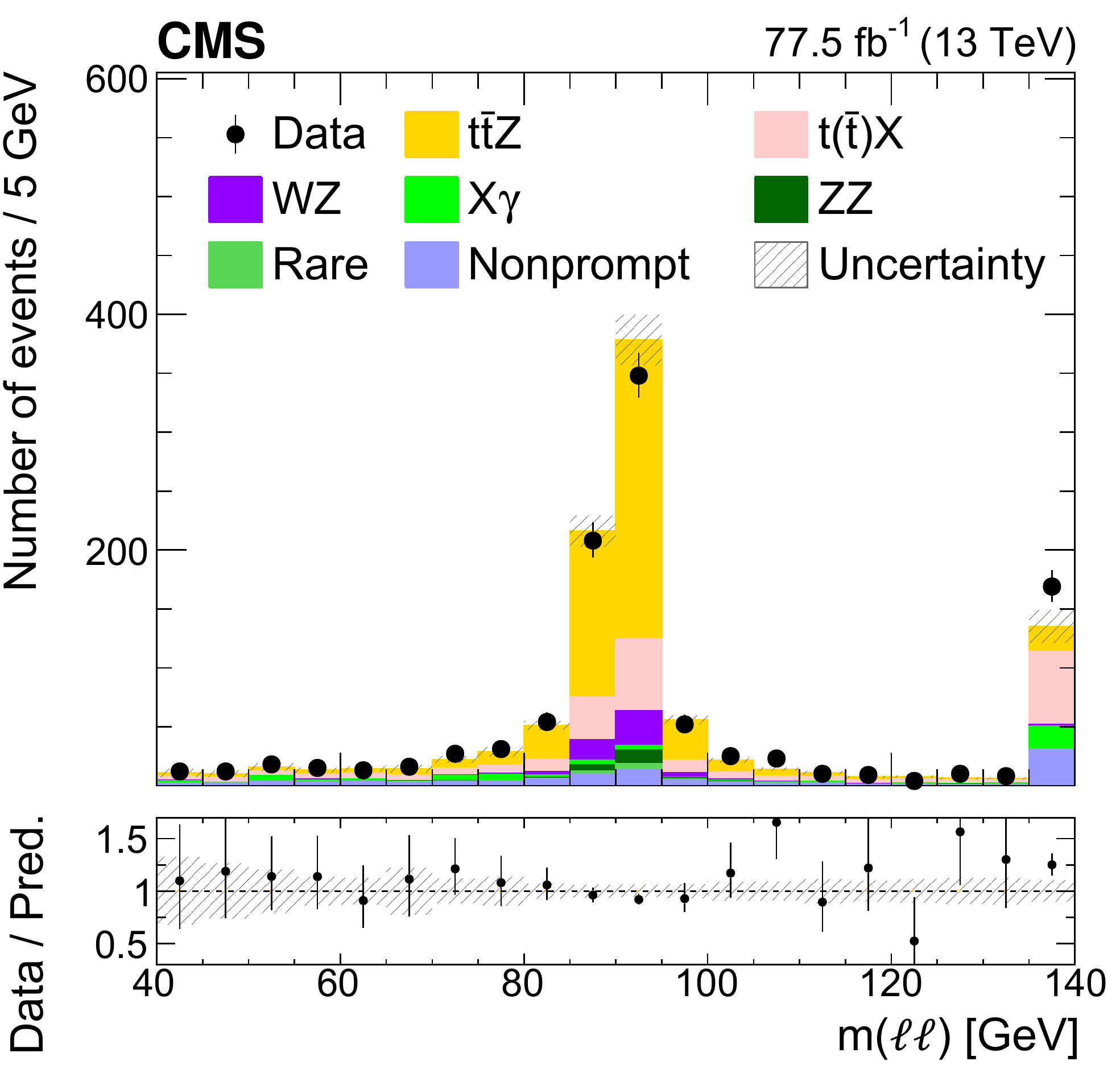}}\\
{\includegraphics[width=0.46\textwidth]{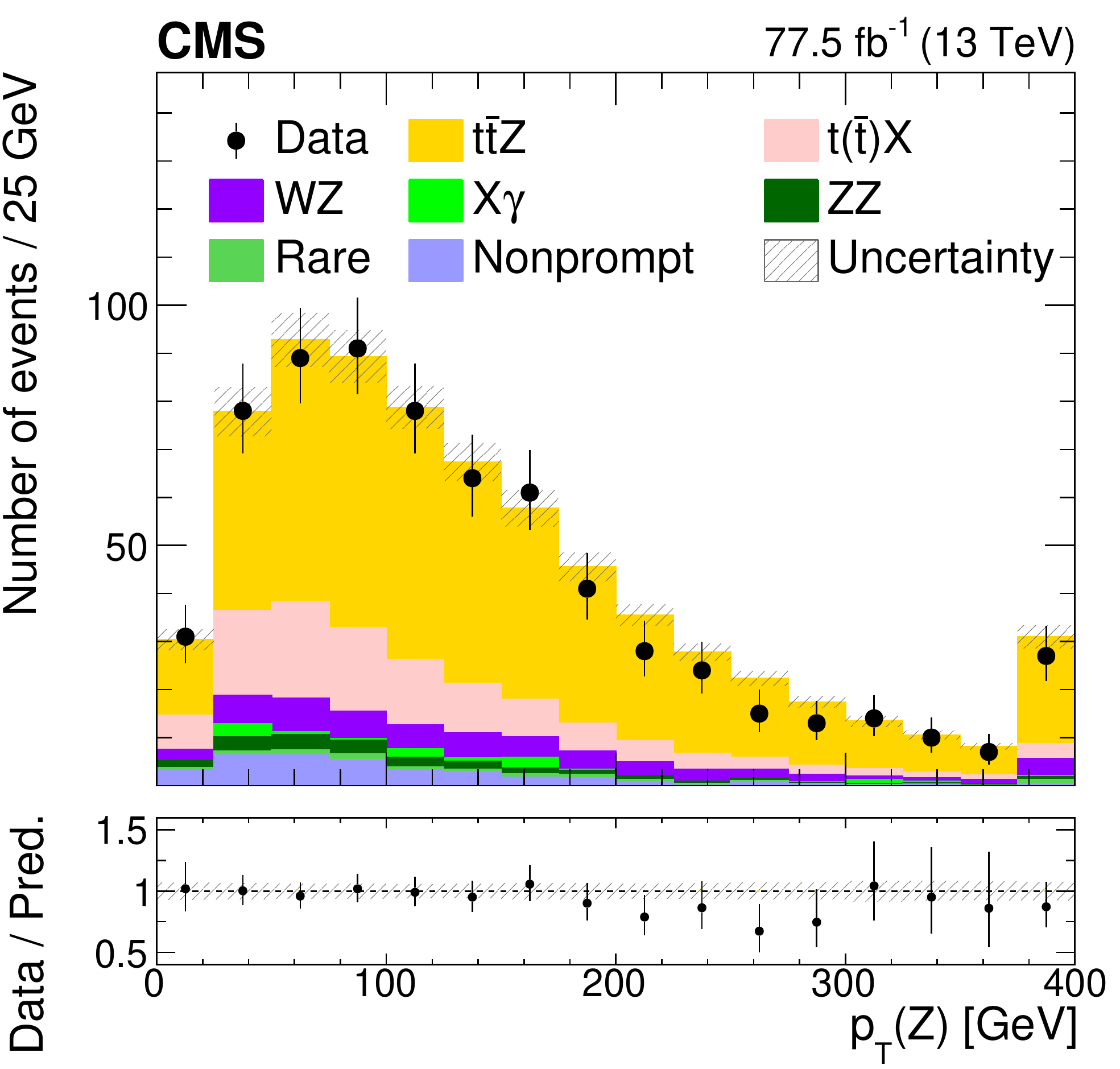}}
{\includegraphics[width=0.46\textwidth]{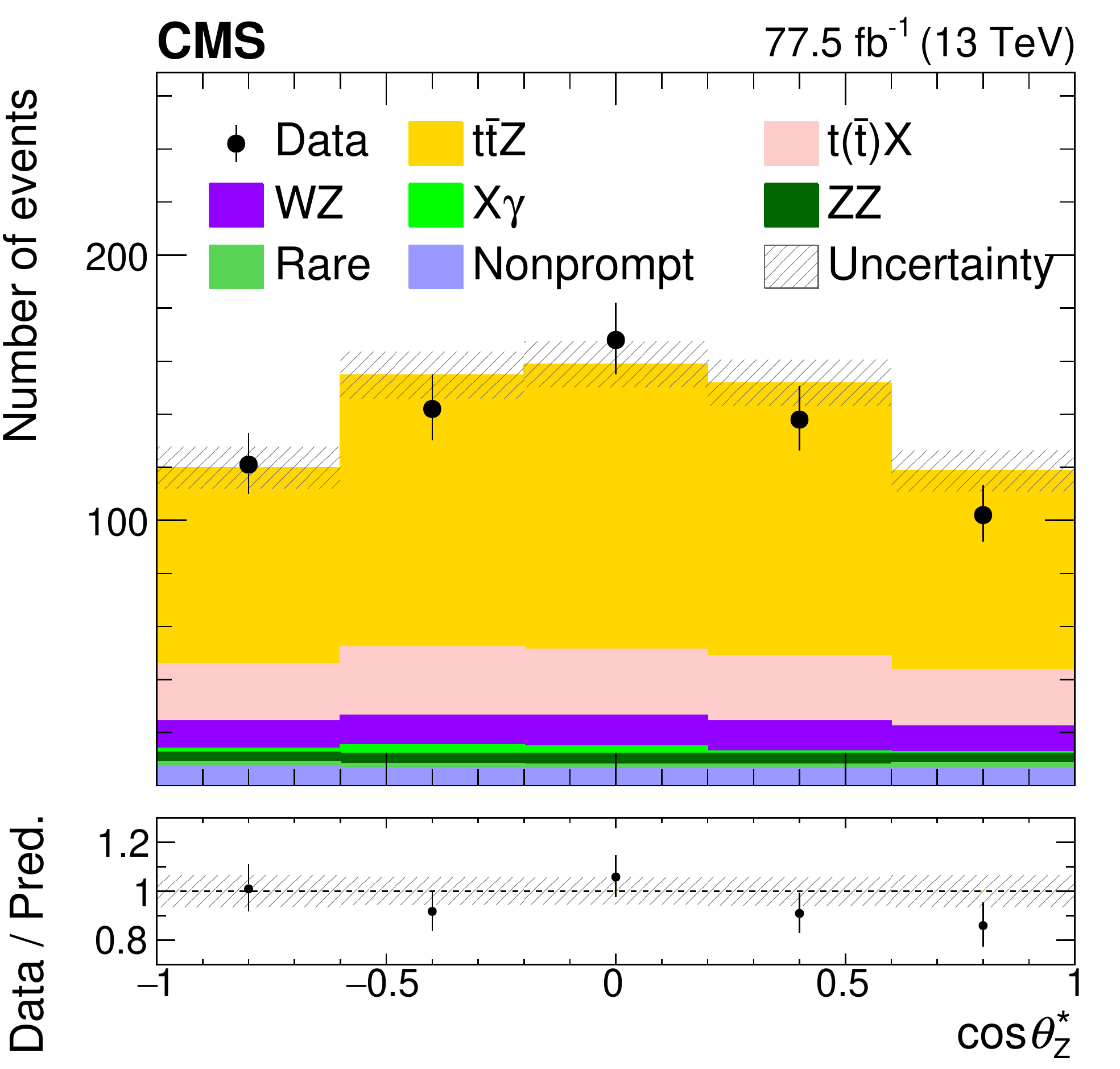}}
  \caption{
Kinematic distributions from a \ttZ signal-enriched subset of events for data (points), compared to the contributions of the signal and background yields from the fit (shaded histograms).
The distributions include the lepton flavor (upper left), number of \cPqb-tagged jets (upper right), jet multiplicity (middle left), dilepton invariant mass \mll (middle right), \pTZ (lower left), and \cosThetaStar (lower right).
The lower panels in each plot give the ratio of the data to the sum of the signal and background from the fit.  The band shows the total uncertainty in the signal and background yields, as obtained from the fit.
}
  \label{fig:ttZ_Kinematics}
\end{figure}

A signal-enriched subset of events is selected by requiring $\Nbjets \geq 1$ and $\Njets \geq 3\,(2)$ for the three (four)-lepton channels. The signal purity is about 65\% for these events.
Figure~\ref{fig:ttZ_Kinematics} shows several kinematic distributions for these signal-enriched events. The sum of the signal and background predictions is found to describe the data within uncertainties. The event yields are listed in Table~\ref{tab:yieldsPureRegion}.

\begin{table}[h!]
\centering\renewcommand{\arraystretch}{1.1}
\topcaption{
The observed number of events for three- and four-lepton events in a signal-enriched sample of events, and the predicted  yields and total uncertainties from the fit for each process.
\label{tab:yieldsPureRegion}}
\begin{tabular}{cccccc}
Process & $\Pgm\Pgm\Pgm(\Pgm)$ & $\Pe\Pgm\Pgm(\Pgm)$ & $\Pe\Pe\Pgm(\Pgm/\Pe)$ & $\Pe\Pe\Pe(\Pe)$ & Total \\[\cmsTabSkip] \hline \noalign{\vskip\cmsTabSkip}
\ttZ  & $ 143 \pm 7.1 $   & $ 122 \pm 6.1 $   & $ 112 \pm 5.5 $   & $ 77 \pm 3.9 $   & $ 455 \pm 22 $  \\
\ttH  & $ 4.1 \pm 0.5 $   & $ 3.5 \pm 0.4 $   & $ 3.3 \pm 0.4 $   & $ 2.1 \pm 0.3 $   & $ 13 \pm 1.6 $  \\
\ttX  & $ 34 \pm 4.2 $   & $ 28 \pm 3.4 $   & $ 24 \pm 2.9 $   & $ 18 \pm 2.3 $   & $ 105 \pm 13 $  \\
\WZ  & $ 18 \pm 4.7 $   & $ 15 \pm 4.2 $   & $ 10 \pm 2.8 $   & $ 11 \pm 3.1 $   & $ 54 \pm 15 $  \\
\Xgamma  & $ 1.8 \pm 1.8 $   & $ 2.1 \pm 2.7 $   & $ 0.6 \pm 0.6 $   & $ 4.6 \pm 1.6 $   & $ 9.0 \pm 3.9 $  \\
\ZZ  & $ 2.8 \pm 0.4 $   & $ 2.7 \pm 0.4 $   & $ 2.5 \pm 0.3 $   & $ 2.2 \pm 0.3 $   & $ 10 \pm 1.3 $  \\
Rare  & $ 2.9 \pm 1.5 $   & $ 2.1 \pm 1.1 $   & $ 1.8 \pm 1.0 $   & $ 1.4 \pm 0.7 $   & $ 8.3 \pm 4.2 $  \\[\cmsTabSkip]
Nonprompt  & $ 6.9 \pm 2.9 $   & $ 11 \pm 4.0 $   & $ 6.9 \pm 2.9 $   & $ 8.5 \pm 3.5 $   & $ 33 \pm 13 $  \\[\cmsTabSkip]
Total & $ 214 \pm 12 $   & $ 187 \pm 12 $   & $ 161 \pm 9.0 $   & $ 125 \pm 8.2 $   & $ 687 \pm 40 $  \\[\cmsTabSkip]
Observed & $ 192 $   & $ 175 $   & $ 152 $   & $ 141 $   & $ 660 $
\end{tabular}
\end{table}

\subsection{Differential cross section measurement}

The differential cross section is measured as a function of \pTZ and \cosThetaStar.
In the simulation, the transverse momentum of the \PZ boson is taken as the final momentum after any QCD and electroweak radiation. The differential cross section is defined in the same phase space as the inclusive cross section reported above, \ie, in the phase space where the top quark pair is produced in association with two leptons with an invariant mass of $70\le\mll\le110$\GeV, corrected for the detector efficiencies and acceptances, as well as for the branching fraction for the \PZ boson decay into a pair of muons or electrons.

The measurement of the differential cross section is performed in a signal-enriched sample of events defined by requiring exactly three identified leptons, $\Nbjets \geq 1$, and $\Njets \geq 3$.
Since the data samples under study are statistically limited, a rather coarse binning in \pTZ and \cosThetaStar is chosen for the differential cross section measurement, with four bins in each distribution.

The cross sections are calculated from the measured event yields corrected for selection and detector effects by subtracting the background and unfolding the resolution effects.
The number of signal events in each bin is determined by subtracting the expected number of background events from the number of events in the data, where the background samples are used without any fit.
The \ttZ \MGvATNLO MC sample is used to construct a response matrix that takes into account both detector response and acceptance corrections. The same corrections, scale factors, and uncertainties as used in the inclusive cross section are applied. Since the resolution of the lepton momenta is good, the fraction of events migrating from one bin to another is extremely small. In all bins, the purity, defined as the fraction of reconstructed events that originate from the same bin, and the stability, defined as the fraction of generated events that are reconstructed in the same bin, are larger than 94\%. Under such conditions, matrix inversion without regularization provides an unbiased and stable method to correct for detector response and acceptance \cite{Cowan:1998ji}. In this analysis, the \texttt{TUnfold} package~\cite{Schmitt:2012kp} is used to obtain the results for the two measured observables.

For each theoretical uncertainty in the signal sample, such as the choice of \muR, \muF, the PDF, and the parton shower, the response matrix is modified and the unfolding procedure is repeated. The uncertainties in the background expectation are accounted for by varying the number of subtracted background events. Experimental uncertainties from the detector response and efficiency, such as the lepton identification, jet energy scale, and \cPqb tagging uncertainties, are applied as a function of the reconstructed observable. For the latter uncertainties, the unfolding is performed using the same response matrix as for the nominal result and varying the input data within their uncertainties. This choice is made in order to minimize possible contributions from numerical effects in the matrix inversion.

\begin{figure}[b!]
  \centering
{\includegraphics[width=0.47\textwidth]{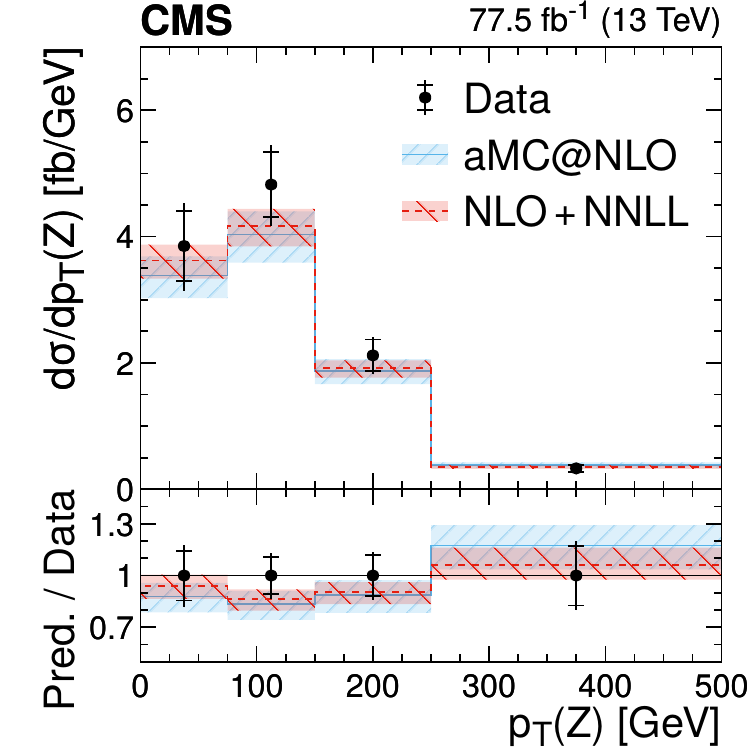}}
\hfill
{\includegraphics[width=0.47\textwidth]{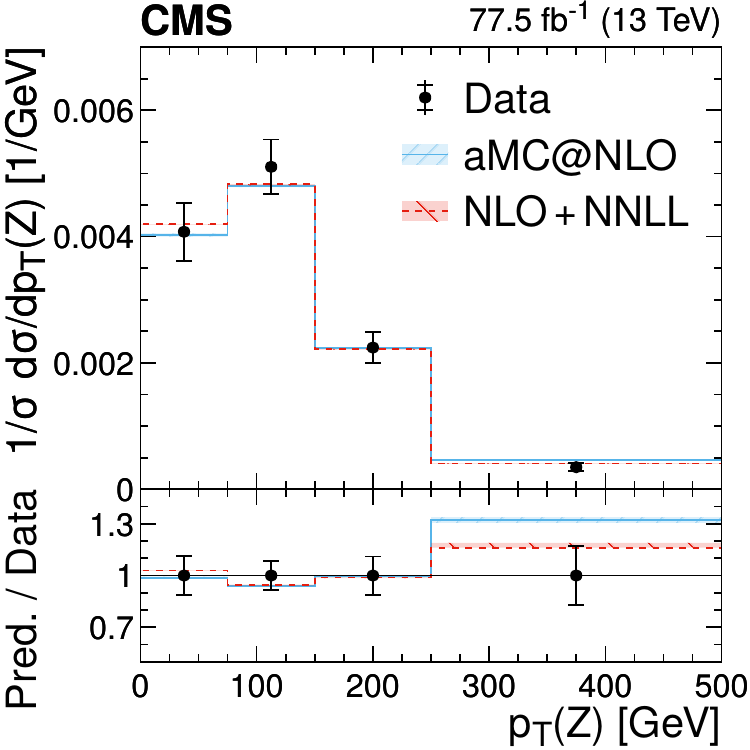}}
\\[\baselineskip]
{\includegraphics[width=0.47\textwidth]{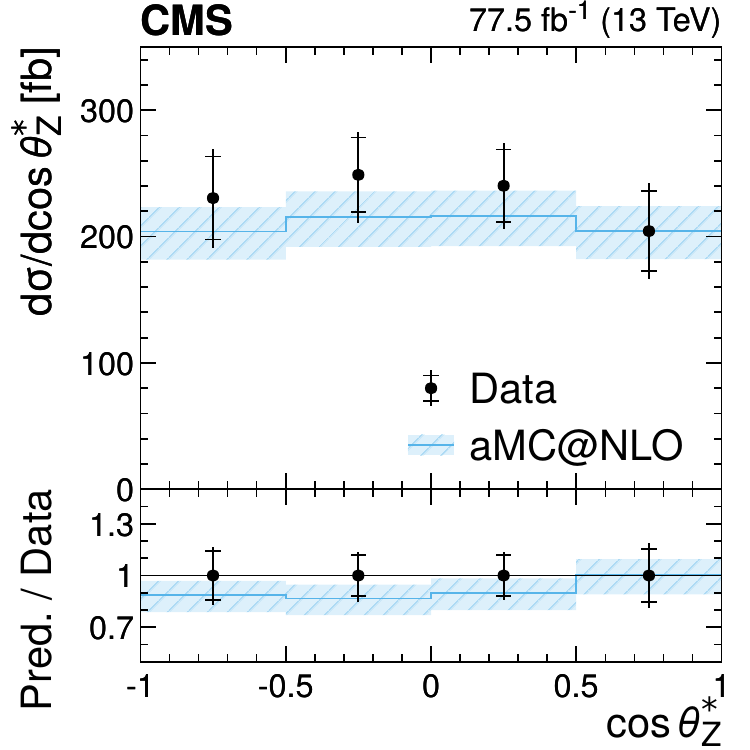}}
\hfill
{\includegraphics[width=0.47\textwidth]{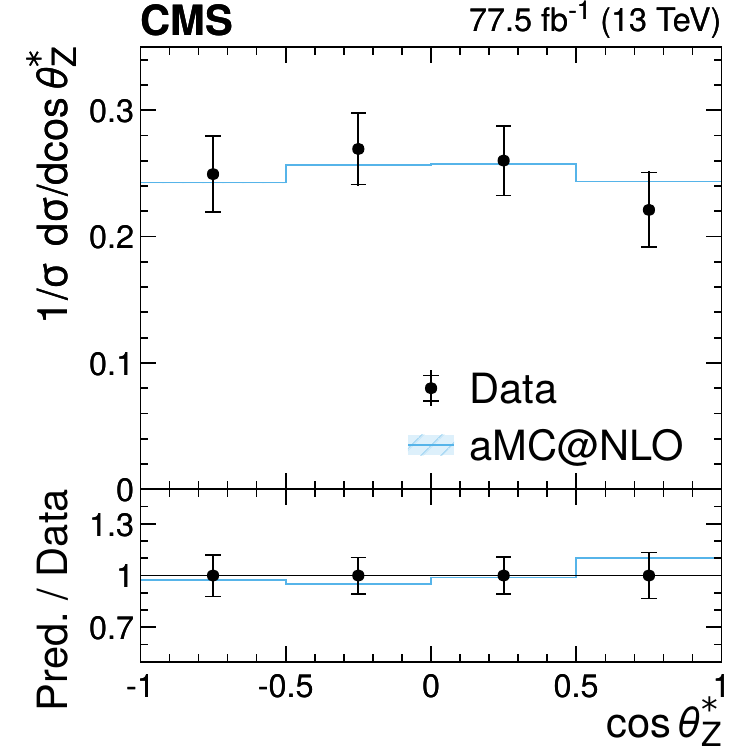}}
  \caption{Measured differential \ttZ production cross sections in the full phase space as a function of the transverse momentum \pTZ of the \PZ boson (upper row) and \cosThetaStar, as defined in the text (lower row). Shown are the absolute (\cmsLeft) and normalized (\cmsRight) cross sections. The data are represented by the points.
The inner (outer) vertical lines indicate the statistical (total) uncertainties.
The solid histogram shows the prediction from the \MGvATNLO MC simulation, and the dashed histogram shows the theory prediction at NLO+NNLL accuracy.
The hatched bands indicate the theoretical uncertainties in the predictions, as defined in the text.
The lower panels display the ratios of the predictions to the measurement.
}
\label{fig:unfolding}
\end{figure}

Figure~\ref{fig:unfolding} left and right show, respectively, the measured absolute and normalized differential cross sections as function of \pTZ and \cosThetaStar, as obtained from the unfolding procedure described above. Also shown is the prediction from the MC generator \MGvATNLO with its uncertainty from scale variations, the PDF choice, and the parton shower~\cite{deFlorian:2016spz,Frixione:2015zaa,Frederix:2018nkq}, as well as a theory prediction at NLO+NNLL accuracy with its uncertainty from scale variations~\cite{Kulesza:2018tqz,Kulesza:2019adl}. Good agreement of the predictions with the measurement is found. The scale variations affect the normalization of the predictions but have negligible impact on their shapes.

\subsection{Search for anomalous couplings and effective field theory interpretation}
\label{sec:EFT}

The role of the top quark in many BSM models~\cite{Hollik:1998vz,Agashe:2006wa,Kagan:2009bn,Ibrahim:2010hv,Ibrahim:2011im,Grojean:2013qca}
makes its interactions, in particular the electroweak gauge couplings,
sensitive probes that can be exploited by interpreting the differential \ttZ cross section in models with modified interactions of the top quark and the \PZ boson.
Extending the earlier analysis~\cite{Sirunyan:2017uzs}, where the inclusive cross section measurement was used, we consider an anomalous coupling Lagrangian~\cite{Rontsch:2014cca}
\begin{align*}
\mathcal{L} = e \overline{u}_\PQt\biggl[ \gamma^{\mu} \bigl(\ConeV + \gamma_5 \ConeA \bigr) + \frac{\mathrm{i} \sigma^{\mu\nu} p_{\nu}} {\mZ}  \bigl(\CtwoV + \mathrm{i} \gamma_5 \CtwoA \bigr)
\biggr] v_\PAQt\,\PZ_{\mu},
\end{align*}
which contains the neutral vector and axial-vector current couplings, $\ConeV$ and $\ConeA$, respectively.
The electroweak magnetic and electric dipole interaction couplings are denoted by $\ConeV$ and $\ConeA$, respectively, and
the four-momentum of the \PZ boson is denoted by $p_\nu$.
In total, there are four real parameters.
The current couplings are exactly predicted by the SM as
\begin{align*}
\ConeV^{\mathrm{SM}}&=\frac{I^\mathrm{f}_{3,\Pq}-2Q_\mathrm{f}\swsq}{2\swcw}= 0.2448\,(52),\\
\ConeA^{\mathrm{SM}}&=\frac{-I^\mathrm{f}_{3,\Pq}}{2\swcw}= -0.6012\,(14),
\end{align*}
where $\thetaw$ is the Weinberg angle,
and $Q_\mathrm{f}$ and $I^\mathrm{f}_{3,\Pq}$ label the charge and the third component of the isospin of the SM fermions, respectively~\cite{Tanabashi:2018oca}.
The dipole moments, moreover, are generated only radiatively in the SM.
Their small numerical values, which are well below $10^{-3}$~\cite{Hollik:1998vz,Bernabeu:1995gs,Czarnecki:1996rx}, therefore allow stringent tests of the SM.
Beyond \pTZ, several observables have been considered that are sensitive to anomalous electroweak interactions of the top quark~\cite{Schulze:2016qas}.
Among them, $\cosThetaStar$ has a high experimental resolution and provides the best discriminating power when compared to a comprehensive set of alternative choices calculated using the reconstructed leptons, jets, and \cPqb-tagged jets.

An alternative interpretation is given in the context of SMEFT in the Warsaw basis~\cite{Grzadkowski:2010es} formed by 59 independent Wilson coefficients of mass dimension 6.
Among them, 15 are important for top quark interactions~\cite{Zhang:2010dr}, which in general have a large impact
on processes other than \ttZ.
Anomalous interactions between the top quark and the gluon (chromomagnetic and chromoelectric dipole moment interactions)
are tightly constrained by the {\ttbar}+jets measurement~\cite{Sirunyan:2018ucr}.
Similarly, the modification of the $\PW\PQt\Pb$ vertex is best constrained by measurements of the \PW~helicity fractions in top quark pair production~\cite{Khachatryan:2016fky}
and in $t$-channel single top quark production~\cite{AguilarSaavedra:2010nx}.
It is thus appropriate to separately consider the operators that induce anomalous interactions of the top quark with the remaining neutral gauge bosons, the \PZ boson and the photon.
In the parametrization adopted here~\cite{AguilarSaavedra:2018nen}, the relevant Wilson coefficients are \ctZ, \ctZI, \cpt, and \cpQM.
The former two induce electroweak dipole moments, while the latter two induce anomalous neutral-current interactions.
These Wilson coefficients, which are combined as
{\allowdisplaybreaks
\begin{align*}
\ctZ  &= \mathrm{Re}\left( -\sinw C_{\PQu\cmsSymbolFace{B}}^{(33)} + \cosw C_{\PQu\PW}^{(33)}\right) \\
\ctZI &= \mathrm{Im}\left( -\sinw C_{\PQu\cmsSymbolFace{B}}^{(33)} + \cosw C_{\PQu\PW}^{(33)}\right) \\
\cpt  &= C_{\phi \PQt} = C_{\phi \PQu}^{(33)}\\
\cpQM &= C_{\phi \cmsSymbolFace{Q}} = C_{\phi \PQq}^{1(33)} - C_{\phi \PQq}^{3(33)},
\end{align*}}
are the main focus of this work. The Wilson coefficients in the Warsaw basis are denoted by $C_{\PQu\cmsSymbolFace{B}}^{(33)}$, $ C_{\PQu\PW}^{(33)}$, $C_{\phi \PQu}^{(33)}$, $C_{\phi \PQq}^{1(33)}$, and $C_{\phi \PQq}^{3(33)}$, as defined in Ref.~\cite{AguilarSaavedra:2018nen}.
The constraints $C_{\phi \PQq}^{3(33)}=0$ and $C_{\PQu\PW}^{(33)}=0$ ensure a SM $\PW\PQt\PQb$ vertex.
Wilson coefficients that are not considered in this work are kept at their SM values and the SMEFT expansion parameter is set to $\Lambda=1\TeV$.

Based on the best expected sensitivity, we choose the following signal regions in the three- and four-lepton channels.
In the three-lepton channel, there are 12 signal regions defined by the four \pTZ thresholds 0, 100, 200, and 400\GeV, and three thresholds on \cosThetaStar at $-1.0$, $-0.6$, and $0.6$.
In the four-lepton channel, the predicted event yields are lower, leading to an optimal choice of only three bins defined in terms of \pTZ with thresholds at 0, 100, and 200\GeV.
The jet multiplicity requirement is relaxed to $\njets\geq 1$.
Next, 12 control regions in the three-lepton channel are defined by requiring $\nbtags=0$ and $\njets\geq 1$, but otherwise reproducing the three-lepton signal selections.
The three-lepton control regions guarantee a pure selection of the main \WZ background.
In order to also constrain the leading \ZZ background of the four-lepton channel, we add three more control regions with $\nbtags\geq 0$ and $\njets\geq 1$ and require that there be two
pairs of opposite-sign same-flavor leptons consistent with the \PZ boson mass in a window of $\pm15\GeV$.
A summary of the signal and control regions is given in Table~\ref{table:srdefinition}.

\begin{table}[!htb]
\centering
\topcaption{Definition of the signal regions (SRs) and control regions (CRs).
For signal regions SR13, SR14, and SR15
and control regions CR13, CR14, and CR15,
there is no requirement on \cosThetaStar.}
\label{table:srdefinition}
\cmsTable{
\begin{tabular}{cccccccc}
\nleptons & \Nbjets & \Njets & $N_{\PZ}$ & \pTZ (\GeVns) & $-1 \leq \cosThetaStar < -0.6 $ & $-0.6 \leq \cosThetaStar < 0.6 $ & $0.6 \leq \cosThetaStar $ \\  \hline
\multirow{4}{*}{3} & \multirow{4}{*}{$\geq$1} & \multirow{4}{*}{$\geq$3} & \multirow{4}{*}{1} & \multirow{1}{*}{0--100} & SR1 & SR2 &  SR3 \\
& & & & \multirow{1}{*}{100--200} & SR4 & SR5 & SR6 \\
& & & & \multirow{1}{*}{200--400} & SR7 & SR8 & SR9 \\
& & & & \multirow{1}{*}{$\geq$400} & SR10 & SR11 & SR12 \\[\cmsTabSkip]
\multirow{3}{*}{4} & \multirow{3}{*}{$\geq$1} & \multirow{3}{*}{$\geq$1} & \multirow{3}{*}{1} & \multirow{1}{*}{0--100} & \multicolumn{3}{c}{SR13} \\
& & & & \multirow{1}{*}{100--200} & \multicolumn{3}{c}{SR14} \\
& & & & \multirow{1}{*}{$\geq$200} & \multicolumn{3}{c}{SR15} \\[\cmsTabSkip]
\multirow{4}{*}{3} & \multirow{4}{*}{0} & \multirow{4}{*}{$\geq$1} & \multirow{4}{*}{1} & \multirow{1}{*}{0--100} & CR1 & CR2 & CR3 \\
& & & & \multirow{1}{*}{100--200} & CR4 & CR5 & CR6 \\
& & & & \multirow{1}{*}{200--400} & CR7 & CR8 & CR9 \\
& & & & \multirow{1}{*}{$\geq$400} & CR10 & CR11 & CR12 \\[\cmsTabSkip]
\multirow{3}{*}{4} & \multirow{3}{*}{$\geq$0} & \multirow{3}{*}{$\geq$1} & \multirow{3}{*}{2} & \multirow{1}{*}{0--100} & \multicolumn{3}{c}{CR13} \\
& & & & \multirow{1}{*}{100--200} & \multicolumn{3}{c}{CR14} \\
& & & & \multirow{1}{*}{$\geq$200} & \multicolumn{3}{c}{CR15} \\ \hline
\end{tabular}
}
\end{table}

The predictions for signal yields with nonzero values of anomalous couplings or Wilson coefficients are obtained by simulating large LO samples in the respective model on a fine
grid in the parameter space, including the SM configuration.
Then, the two-dimensional (2D) generator-level distributions of \pTZ and \cosThetaStar for the BSM and the SM parameter points are used to define the reweighting of the nominal NLO \ttZ sample.
The result of the reweighting procedure is tested on a coarse grid in BSM parameter space, where BSM samples are produced and reconstructed.
The differences between the full event reconstruction and the reweighting procedure are found to be negligible for all distributions considered in this work.
The theoretical uncertainties in the predicted BSM yields are scaled accordingly.

From the predicted yields and the uncertainties, we construct a binned likelihood function $L(\theta)$ as a product of Poisson probabilities, where $\theta$ labels the set of nuisance parameters.
The test statistic is the profile likelihood ratio $q=-2\ln(L(\hat{\theta}, \vec C)/L(\hat{\theta}_{\text{max}}))$ where $\hat{\theta}$ is the set of nuisance parameters maximizing the likelihood function at a BSM point defined by the Wilson coefficients collectively denoted by $\vec C$. In the denominator, $\hat{\theta}_\text{max}$ maximizes the likelihood function in the BSM parameter plane.

\begin{figure}[!htbp]
\centering
\includegraphics[width=.90\textwidth]{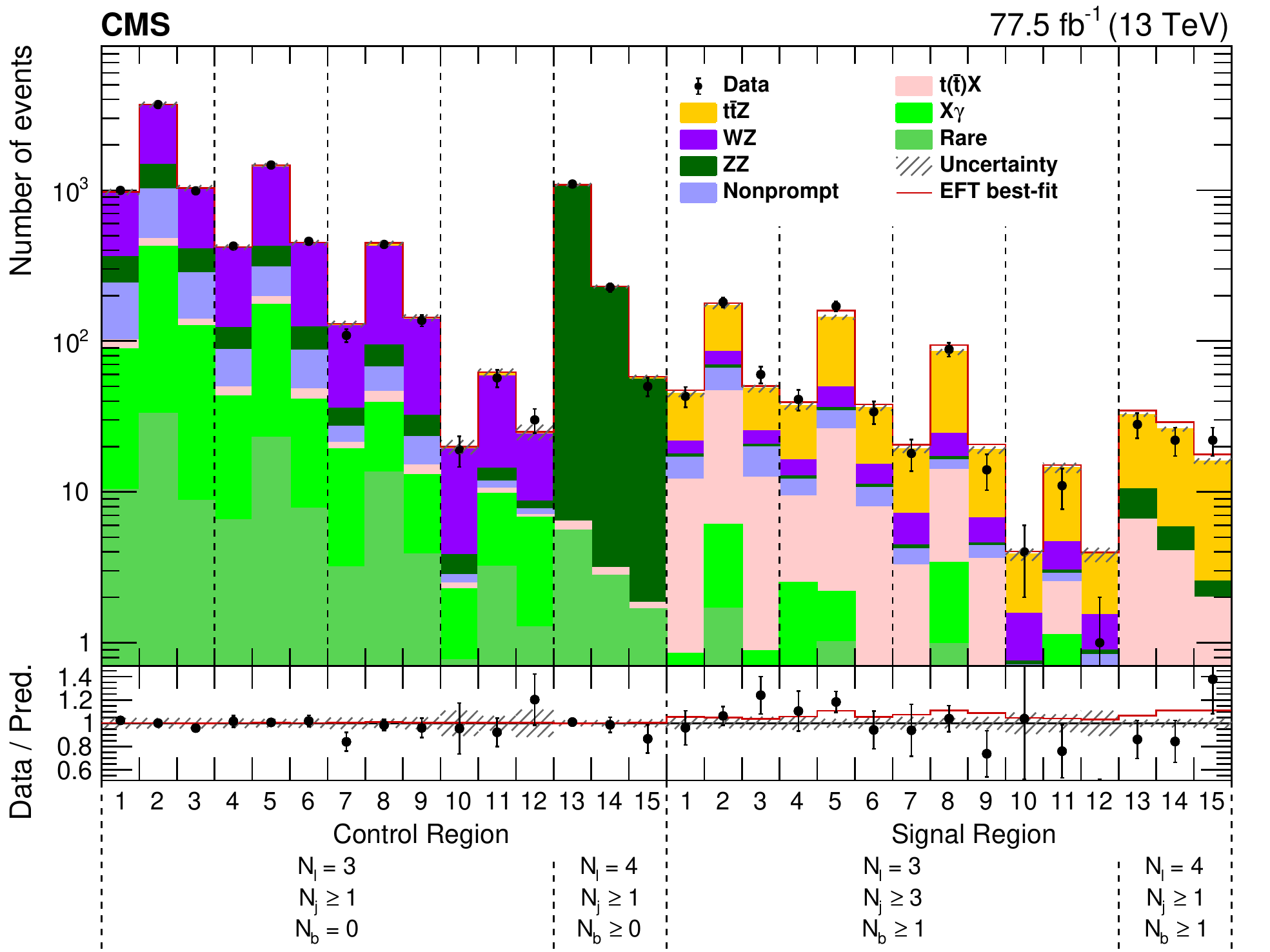}
\caption{
The observed (points) and predicted (shaded histograms) post-fit yields for the combined 2016 and 2017 data sets in the control and signal regions.
In the $\nleptons=3$ control and signal regions (bins 1--12), each of the four \pTZ categories is further split into three \cosThetaStar bins.
The horizontal bars on the points give the statistical uncertainties in the data.
The lower panel displays the ratio of the data to the predictions and the hatched regions show the total uncertainty.
The solid line shows the best-fit prediction from the SMEFT fit.
}
\label{figures:regions_EFT}
\end{figure}

\begin{figure}[!htbp]
\centering
\includegraphics[width=.49\textwidth]{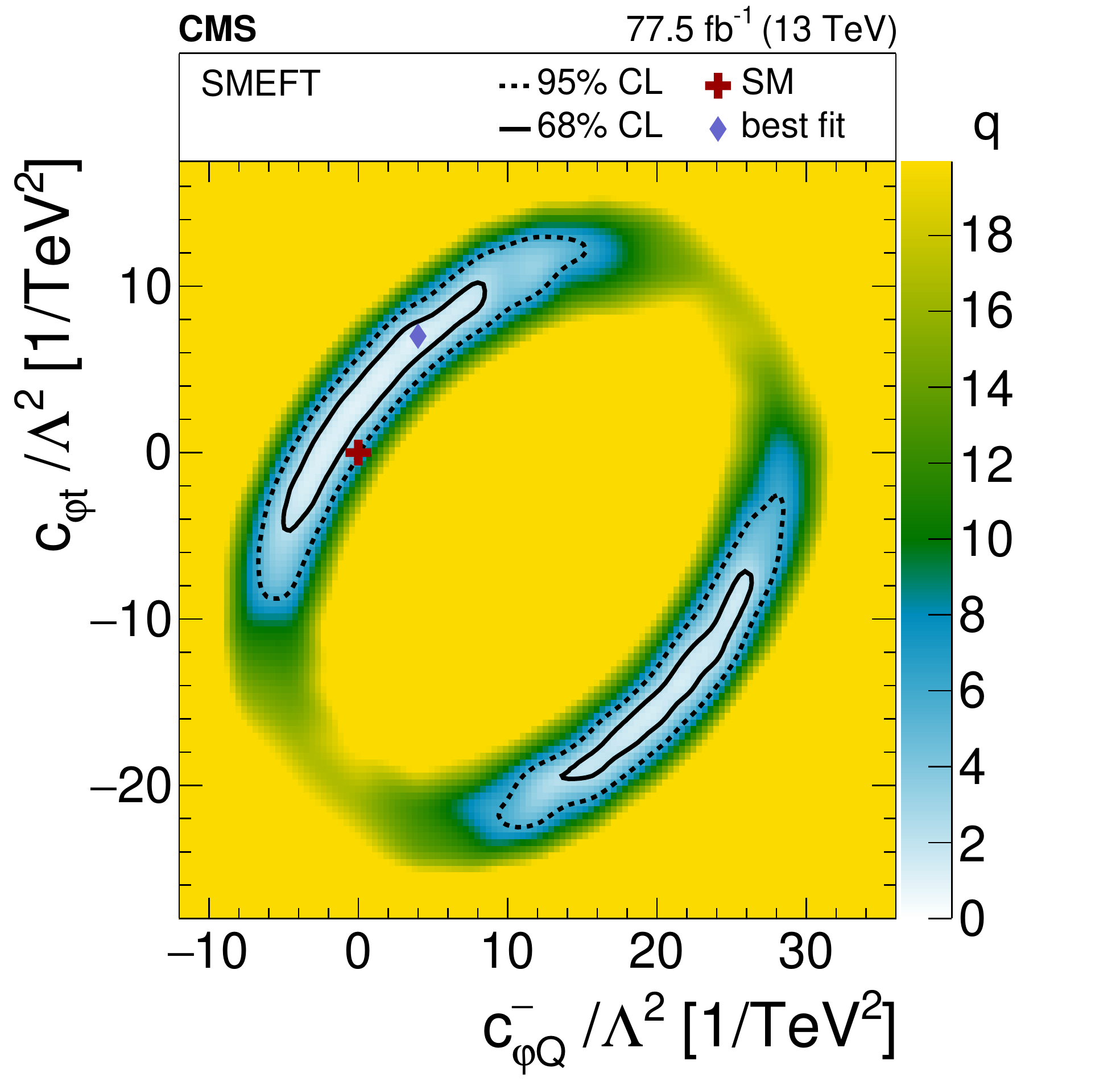}
\includegraphics[width=.49\textwidth]{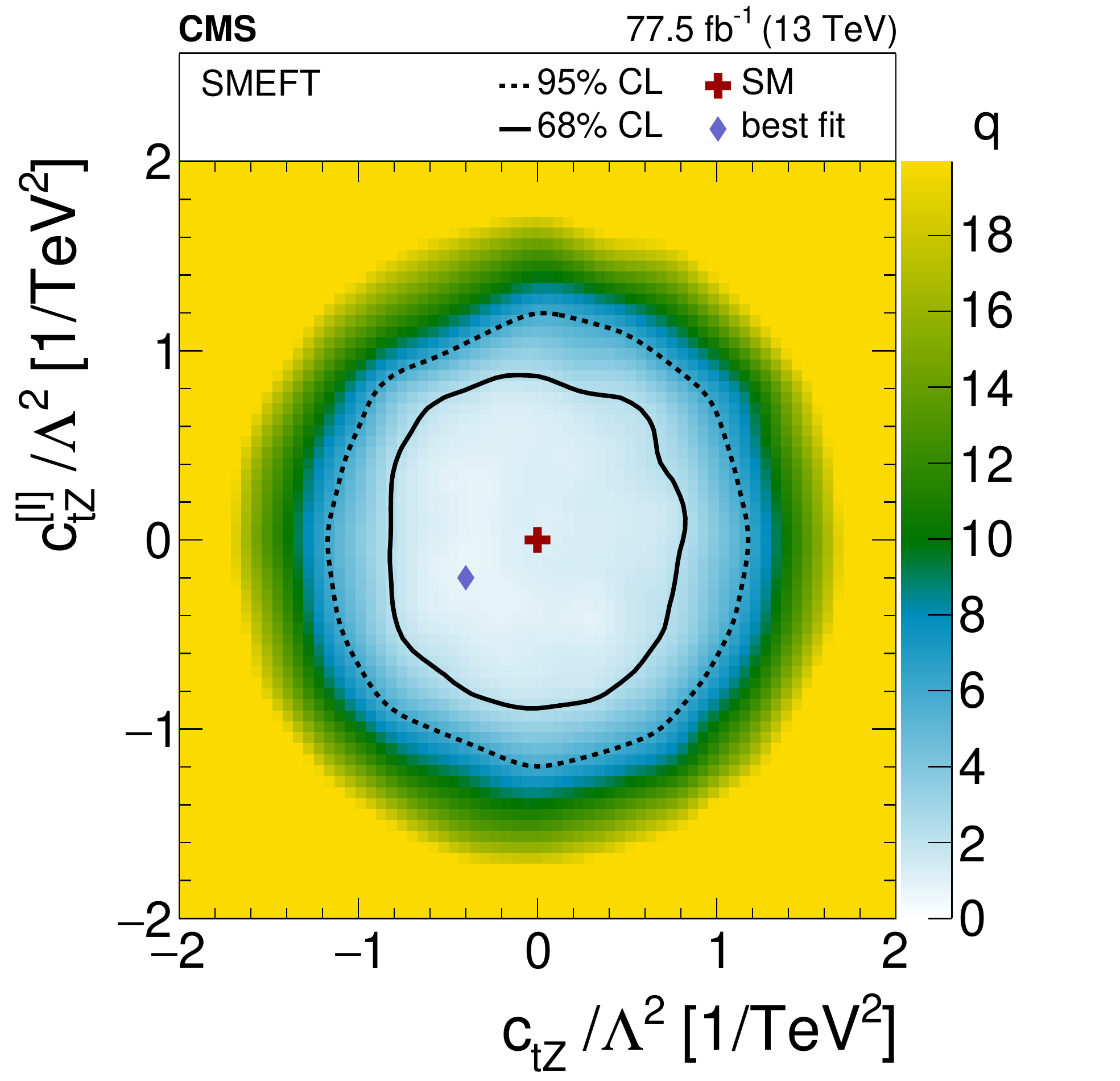}\\
\caption{
Results of scans in two 2D planes for the SMEFT interpretation.
The shading quantified by the gray scale on the right reflects the negative log-likelihood ratio $q$ with respect to the best-fit value, designated by the diamond.
The solid and dashed lines indicate the 68 and 95\% \CL contours from the fit, respectively.
The cross shows the SM prediction.
}
\label{figures:EFT_results}
\end{figure}

\begin{figure}[!htbp]
\centering
\includegraphics[width=.49\textwidth]{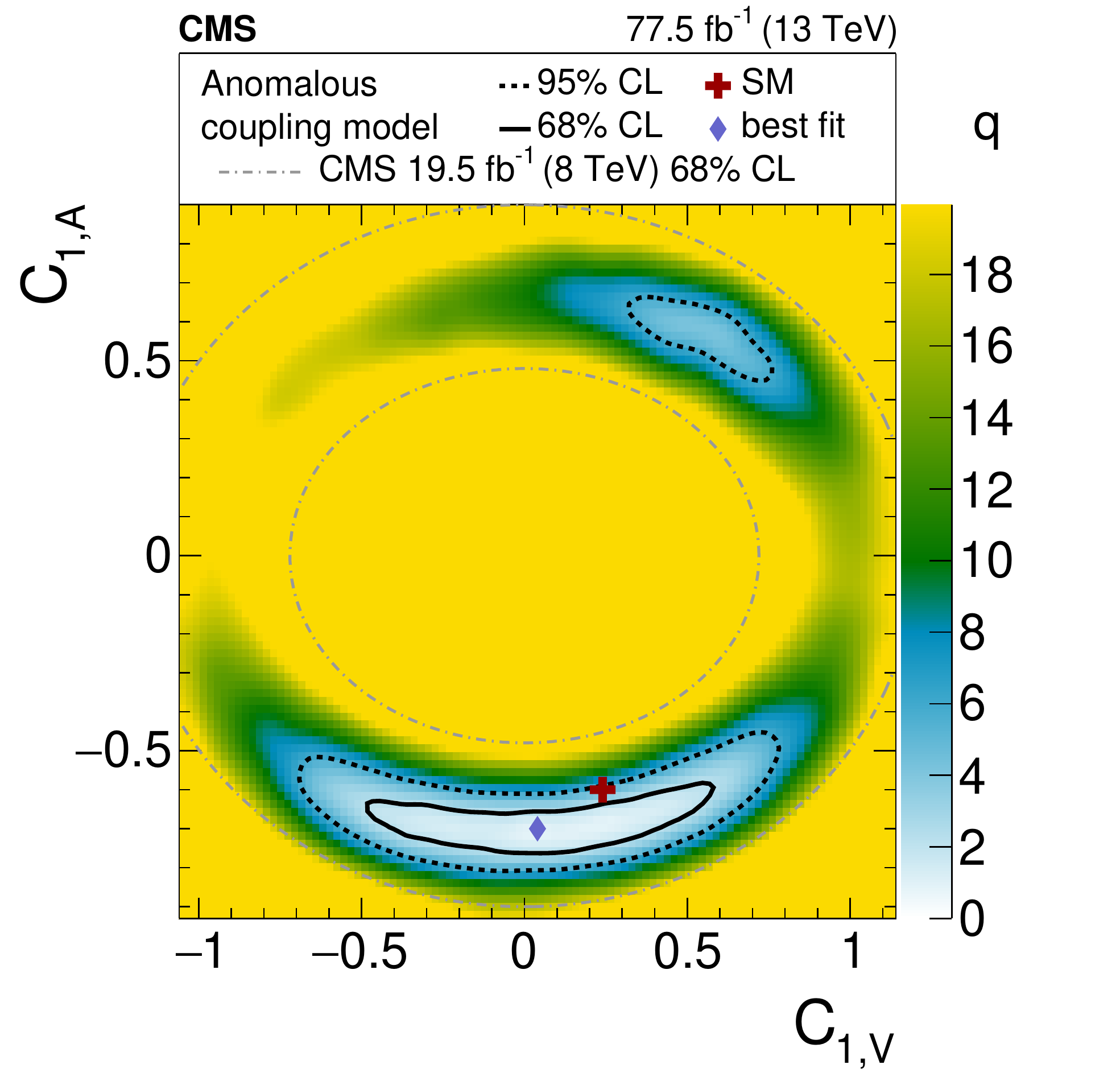}
\includegraphics[width=.49\textwidth]{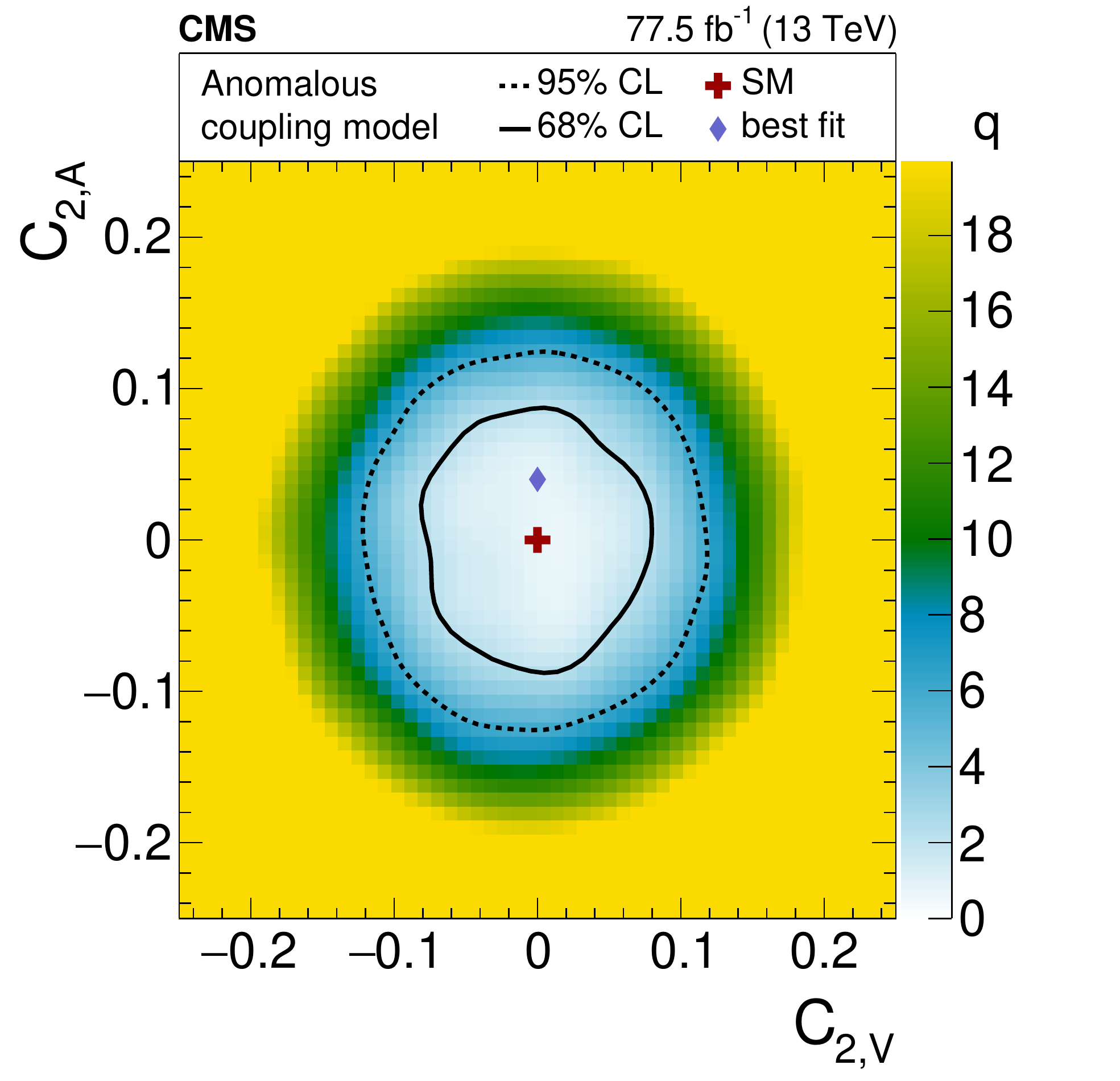}\\
\caption{
Results of scans in the axial-vector and vector current coupling plane (\cmsLeft) and the electroweak dipole moment plane (\cmsRight).
The shading quantified by the gray scale on the right of each plot reflects the log-likelihood ratio $q$ with respect to the best-fit value, designated by the diamond.
The solid and dashed lines indicate the 68 and 95\% \CL contours from the fit, respectively.
The cross shows the SM prediction.
The area between the dot-dashed ellipses in the axial-vector and vector current coupling plane corresponds to the observed 68\% \CL area from the previous CMS result~\cite{Khachatryan:2015sha}.
}
\label{figures:BSM_results}
\end{figure}

Figure~\ref{figures:regions_EFT} shows the best-fit result in the plane spanned by \cpt and \cpQM
using the regions in Table~\ref{table:srdefinition}.
Figure~\ref{figures:EFT_results} displays the log-likelihood scan in the 2D planes spanned by \cpt and \cpQM, as well as \ctZ and \ctZI.
Consistent with the measurement of the cross section, the SM value is close to the contour in 2D at 95\% confidence level (\CL) for modified vector and axial-vector current couplings.
Models with nonzero electroweak dipole moments predict a harder \pTZ spectrum that is not observed in data.
A systematic uncertainty from an effect of nonzero Wilson coefficients on the background prediction, in particular of the \tZq process amounting to a total of less than 8.5\% in the most sensitive bins,
was checked to have a negligible impact.
The SM prediction is within the 68\% confidence interval of the best-fit value of the \ctZ and \ctZI coefficients.
Figure~\ref{figures:BSM_results} shows the complementary scan in the 2D plane spanned by the anomalous current interactions $\ConeV$ and $\ConeA$, as well as the anomalous dipole interactions $\CtwoV$ and $\CtwoA$. In both cases, the SM predictions are consistent with the measurements.

\begin{figure}[!htbp]
\centering
\includegraphics[width=.40\textwidth]{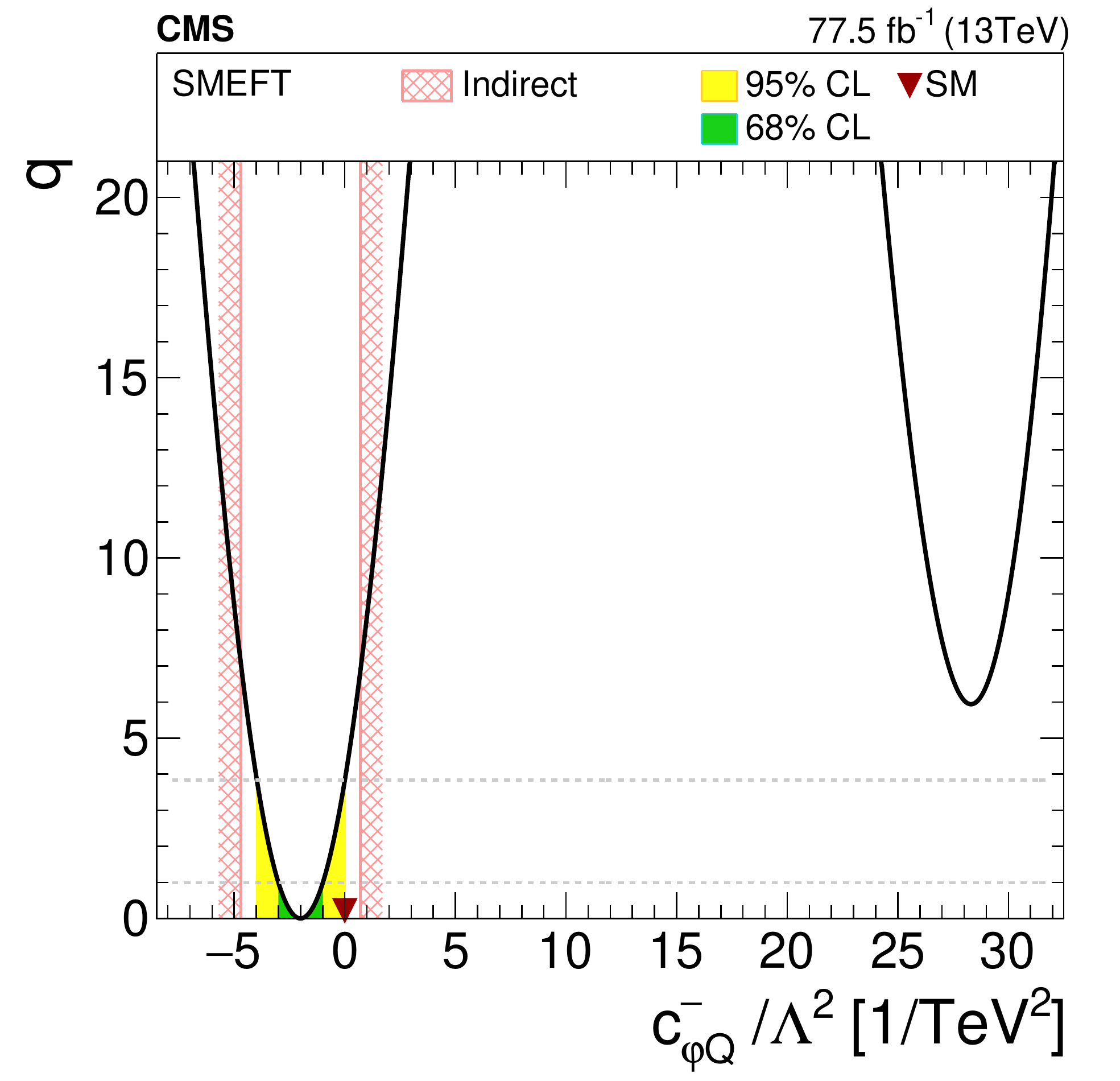}
\includegraphics[width=.40\textwidth]{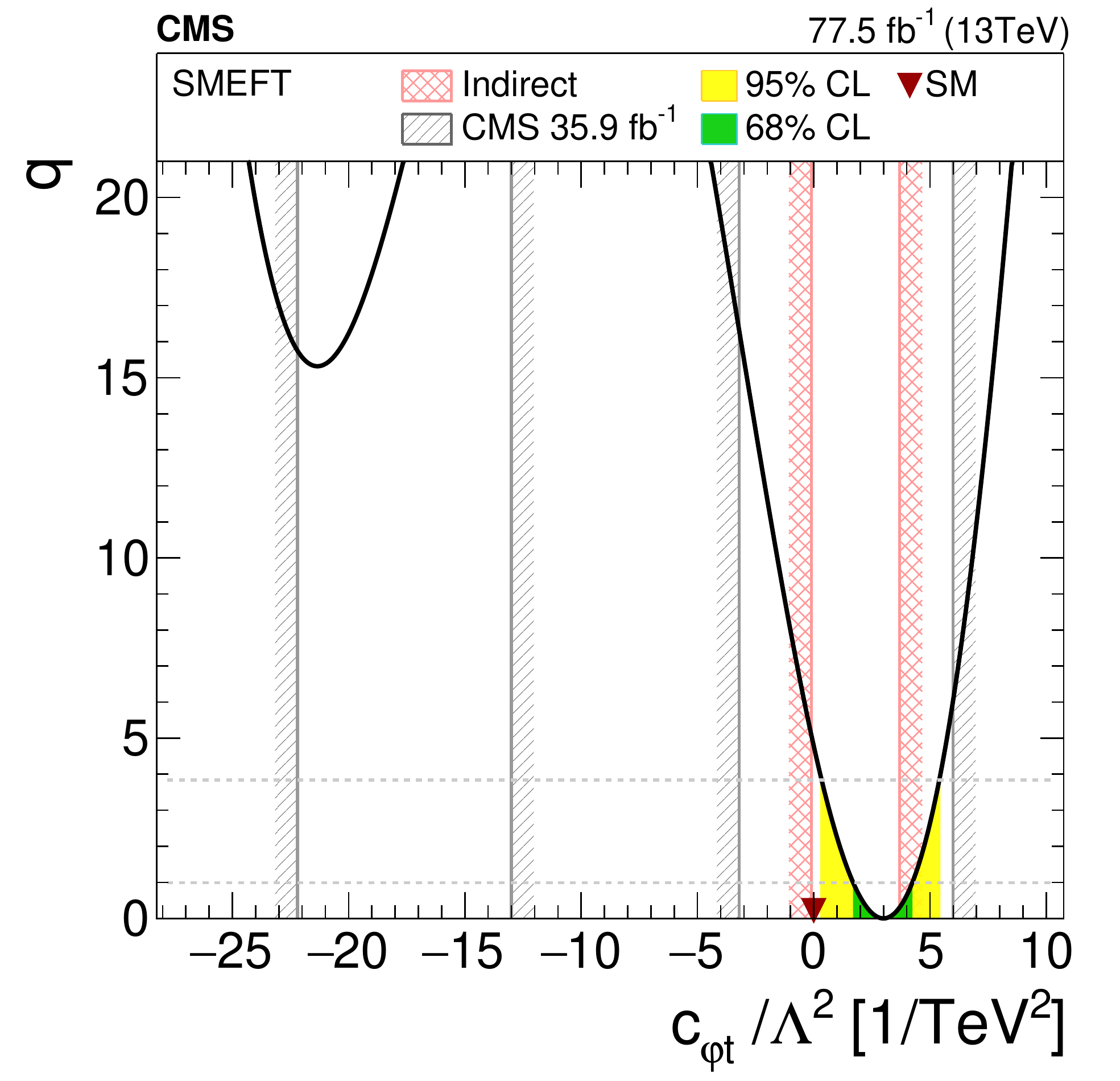}\\
\includegraphics[width=.40\textwidth]{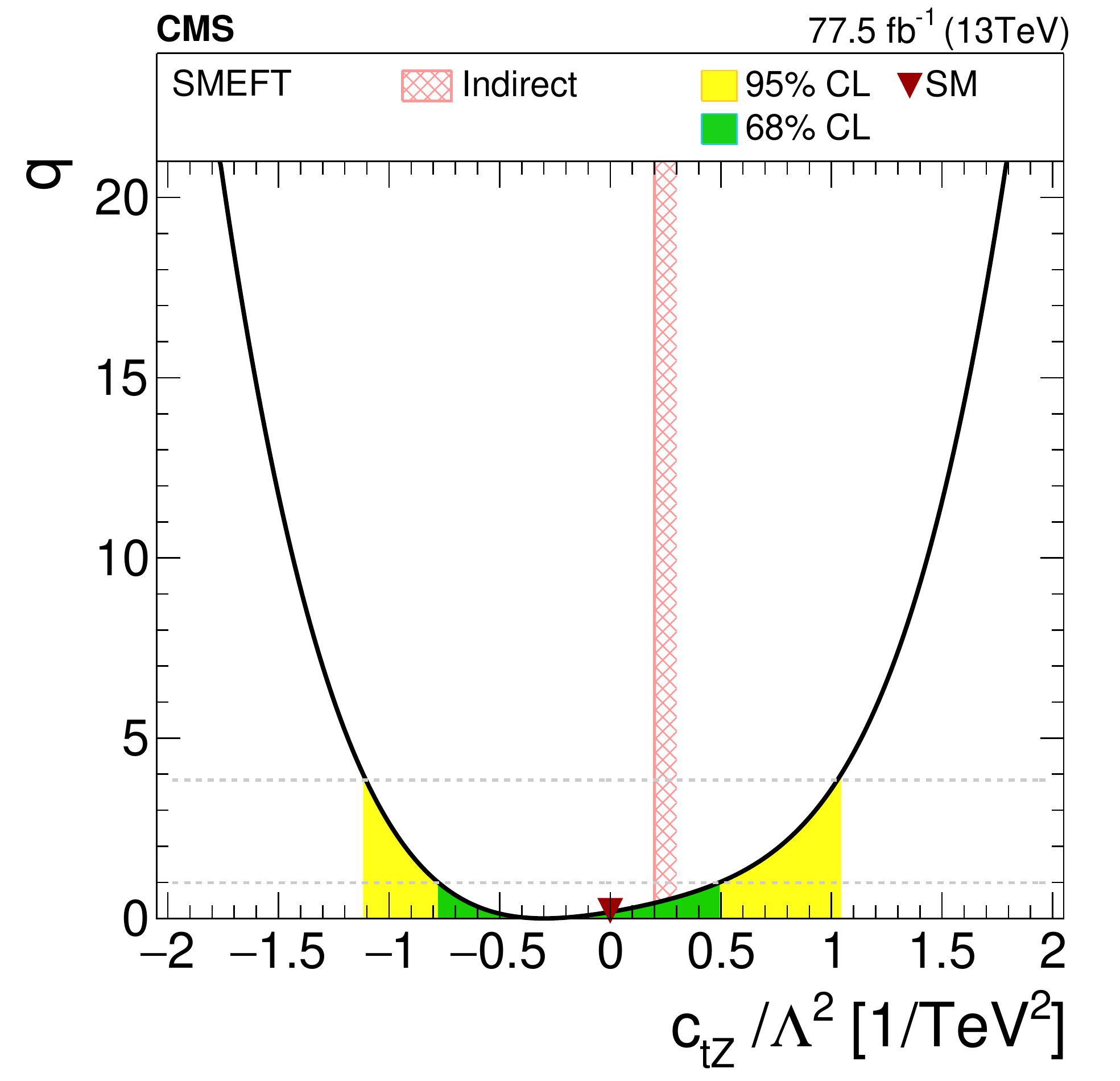}
\includegraphics[width=.40\textwidth]{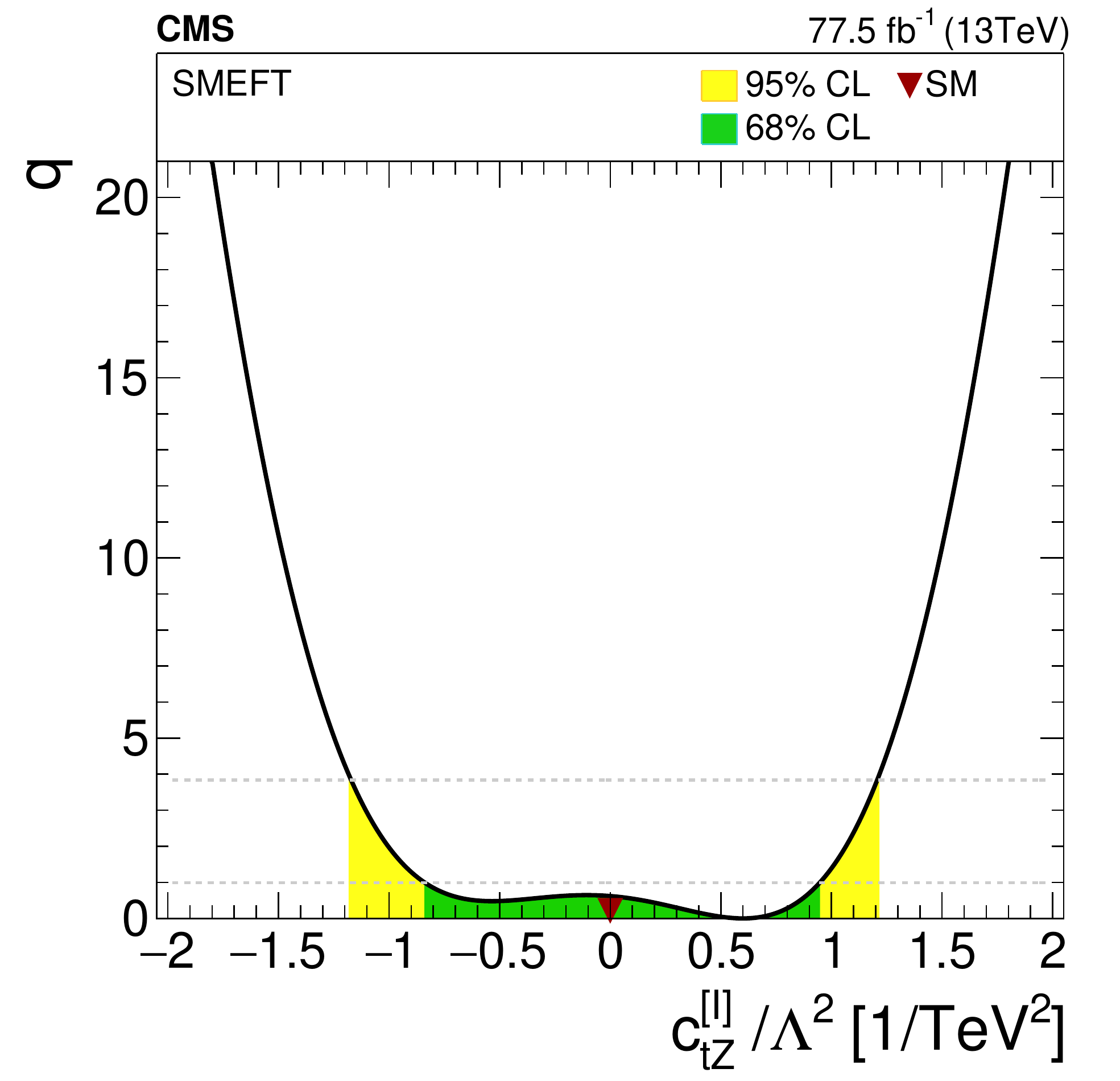}\\
\caption{
1D scans of two Wilson coefficients, with the value of the other Wilson coefficients set to zero.
The shaded areas correspond to the 68 and 95\% \CL intervals around the best fit value, respectively. The downward triangle indicates the SM value.
Previously excluded regions at 95\% \CL \cite{Sirunyan:2017uzs} (if available) are indicated by the hatched band.
Indirect constraints from Ref. \cite{Zhang:2012cd} are shown as a cross-hatched band.
}
\label{figures:EFT_results1D}
\end{figure}

\begin{figure}[!htbp]
\centering
\includegraphics[width=.40\textwidth]{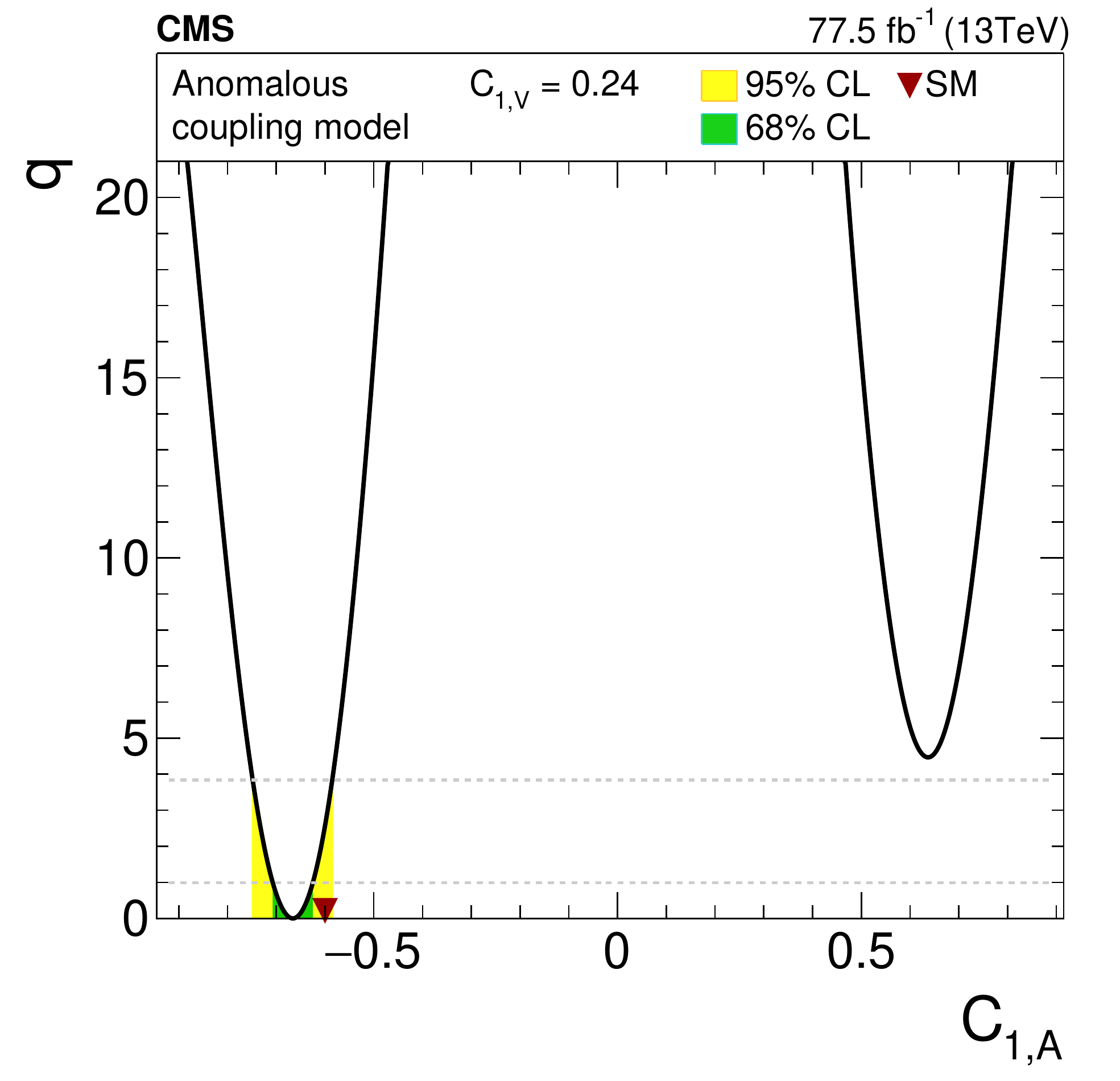}
\includegraphics[width=.40\textwidth]{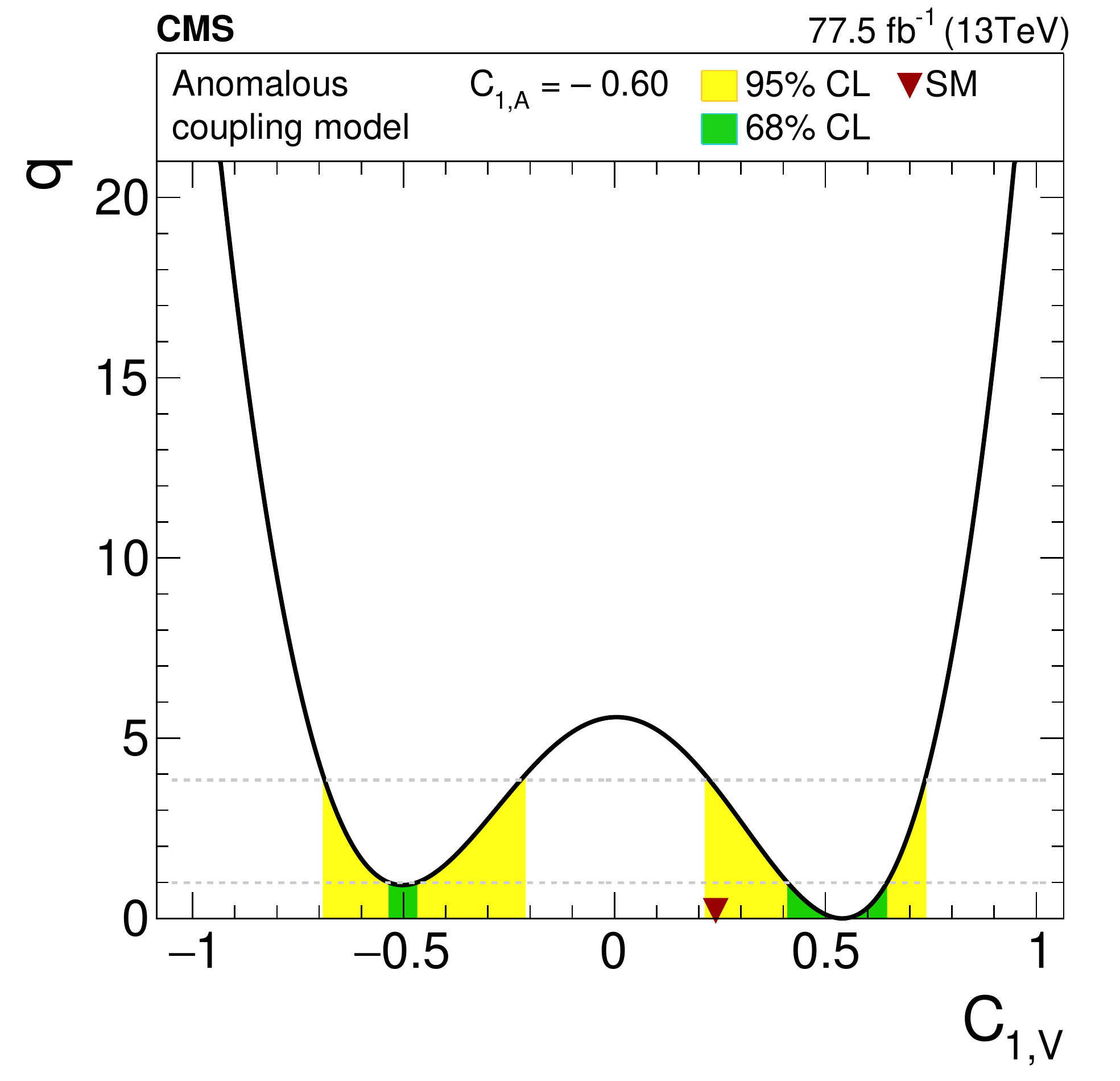}\\
\includegraphics[width=.40\textwidth]{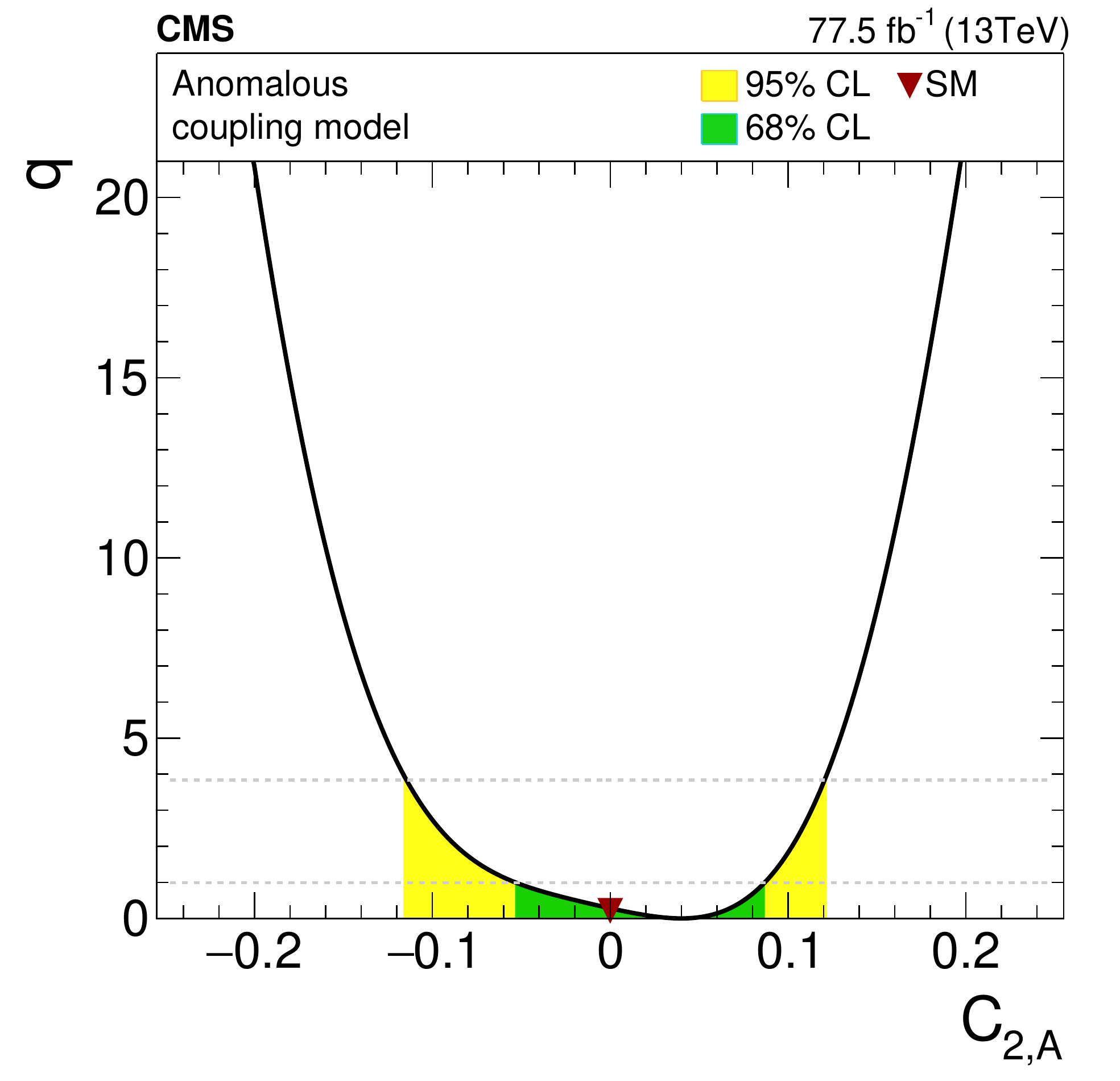}
\includegraphics[width=.40\textwidth]{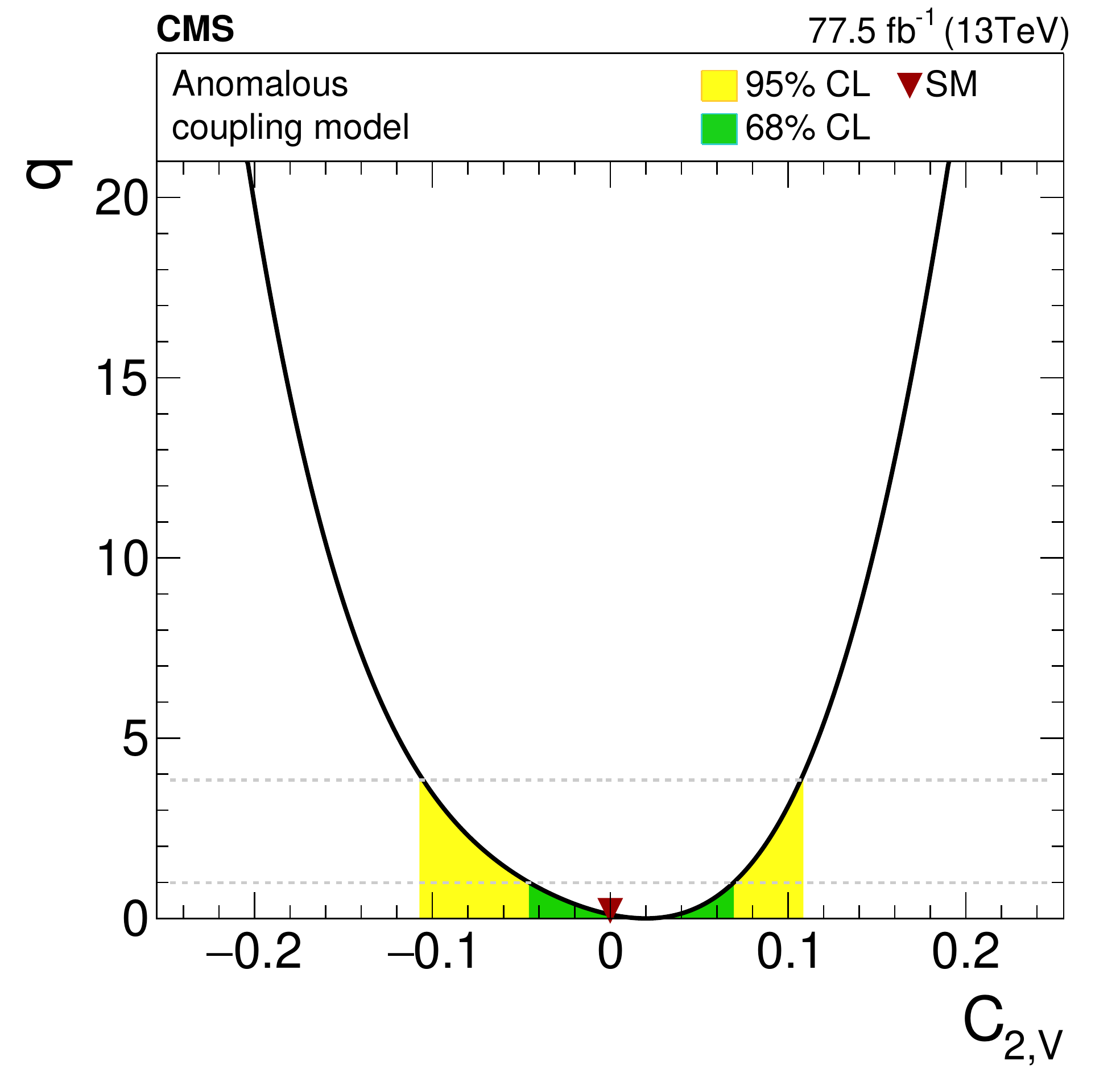}\\
\caption{
Log-likelihood ratios for 1D scans of anomalous couplings.
For the scan of $\ConeA$ (upper \cmsLeft), $\ConeV$ was set to the SM value of 0.24, and for the scan of $\ConeV$ (upper \cmsRight), $\ConeA$ was set to the SM value of $-0.60$.
For the scans of $\CtwoA$ (lower \cmsLeft) and $\CtwoV$ (lower \cmsRight), which correspond to the top quark electric and magnetic dipole moments, respectively, both $\ConeV$ and $\ConeA$ are set to the SM values.
The shaded areas correspond to the 68 and 95\% \CL intervals around the best-fit value, respectively.
The downward triangle indicates the SM value.
}
\label{figures:BSM_results1D}
\end{figure}

\begin{figure}[!htb]
\centering
\includegraphics[width=.60\textwidth]{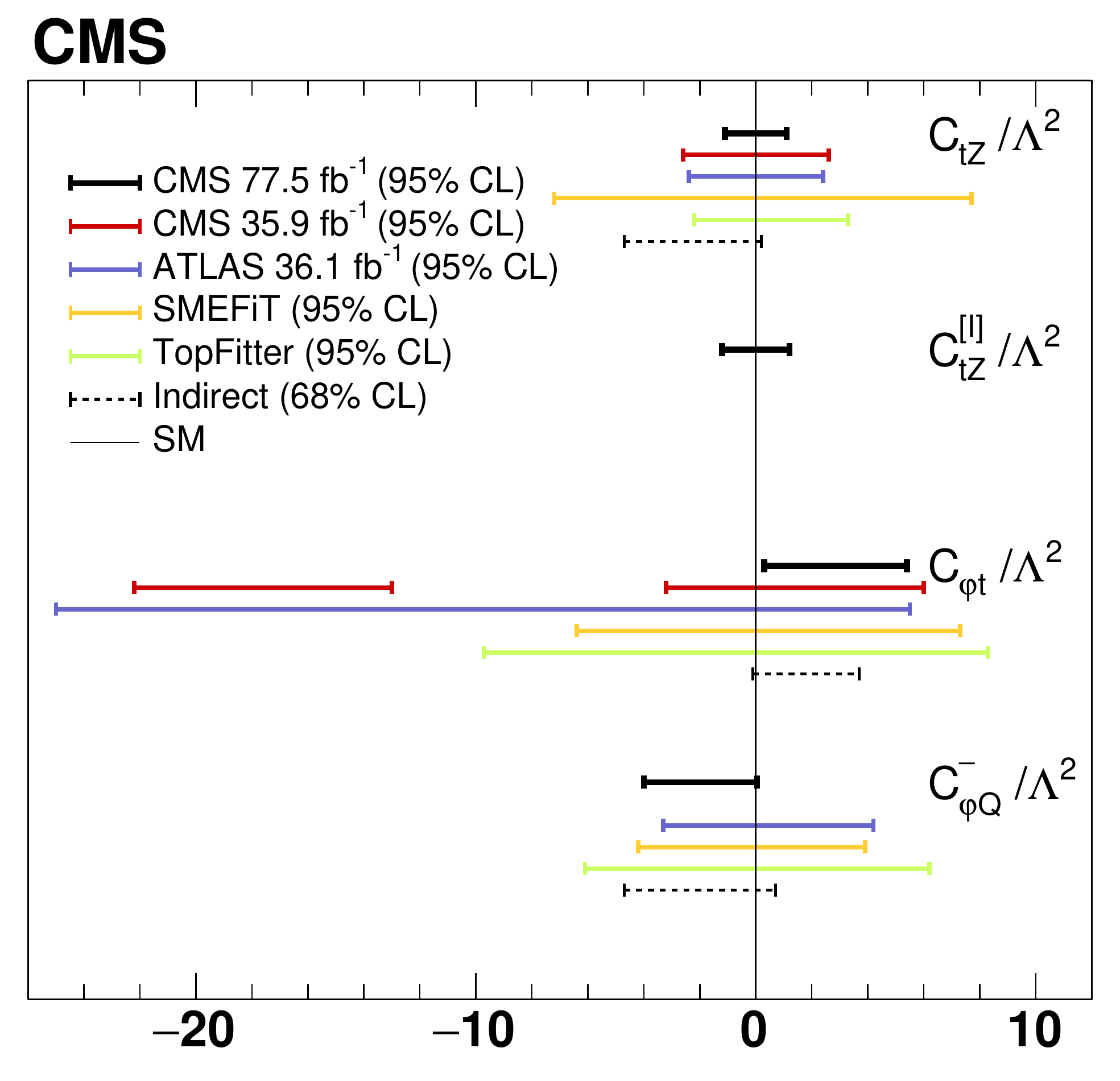}
\caption{
The observed 95\% \CL intervals for the Wilson coefficients from this measurement, the previous CMS result based on the inclusive \ttZ cross section measurement~\cite{Sirunyan:2017uzs}, and
the most recent ATLAS result~\cite{Aaboud:2019njj}.
The direct limits within the SMEFiT framework~\cite{Hartland:2019bjb} and from the TopFitter Collaboration~\cite{Buckley:2015lku},
and the 68\% \CL indirect limits from electroweak data are also shown~\cite{Zhang:2012cd}.
The vertical line displays the SM prediction.
}
\label{figures:EFT_summary}
\end{figure}

Finally, Figs.~\ref{figures:EFT_results1D} and \ref{figures:BSM_results1D} display the one-dimensional (1D) scans, where in each plot, all other coupling parameters are set to their SM values.
The corresponding 1D confidence intervals at 68 and  95\% \CL are listed in Table~\ref{table:limits} and are the most stringent direct constraints to date.
A comparison of the observed 95\% confidence intervals with earlier measurements is shown in Fig.~\ref{figures:EFT_summary}, together
with direct limits obtained within the SMEFiT framework~\cite{Hartland:2019bjb} and by the TopFitter Collaboration~\cite{Buckley:2015lku}.

\begin{table}[!htbp]
\centering
\topcaption{Expected and observed 68 and 95\% \CL intervals from this measurement for the listed Wilson coefficients.
The expected and observed 95\% \CL intervals from a previous CMS measurement \cite{Sirunyan:2017uzs} and indirect 68\% \CL constraints from precision electroweak data \cite{Zhang:2012cd} are shown for comparison.
}
\label{table:limits}
\cmsTable{
\begin{tabular}{cccccccc}
Coefficient & \multicolumn{2}{c}{Expected} & \multicolumn{2}{c}{Observed} & \multicolumn{2}{c}{Previous CMS constraints} & Indirect constraints  \\
& 68\% \CL & 95\% \CL & 68\% \CL & 95\% \CL & Exp. 95\% \CL & Obs. 95\% \CL & 68\% \CL \\ \hline \noalign{\vskip\cmsTabSkip}
$\ctZ / \Lambda^2$ & $[-0.7,~0.7]$ & $[-1.1,~1.1]$ & $[-0.8,~0.5]$ & $[-1.1,~1.1]$ & $[-2.0,~2.0]$ & $[-2.6,~2.6]$ & $[-4.7,~0.2]$ \\[\cmsTabSkip]
$\ctZI / \Lambda^2 $ & $[-0.7,~0.7]$ & $[-1.1,~1.1]$ & $[-0.8,~1.0]$ & $[-1.2,~1.2]$ & \NA  & \NA & \NA \\[\cmsTabSkip]
\multirow{2}{*}{$\cpt / \Lambda^2 $}    & \multirow{2}{*}{$[-1.6,~1.4]$}  & \multirow{2}{*}{$[-3.4,~2.8]$}  & \multirow{2}{*}{$[1.7,~4.2]$} & \multirow{2}{*}{$[0.3,~5.4]$} & \multirow{2}{*}{$[-20.2,~4.0]$} & $[-22.2,~-13.0]$ & \multirow{2}{*}{$[-0.1,~3.7]$} \\
 &   &   &   &   &   & $[-3.2,~6.0]$  &  \\[\cmsTabSkip]
$\cpQM / \Lambda^2 $  & $[-1.1,~1.1]$ & $[-2.1,~2.2]$ & $[-3.0,~-1.0]$ & $[-4.0,~0.0]$ & \NA & \NA & $[-4.7,~0.7]$
\end{tabular}
}
\end{table}

\section{Summary}
\label{sec:Conclusions}
A measurement of top quark pair production in association with a \PZ boson using a data sample of proton-proton collisions at $\sqrt{s}=13\TeV$, corresponding to an integrated luminosity of 77.5\fbinv, collected with the CMS detector at the LHC has been presented.
The analysis was performed in the three- and four-lepton final states using analysis categories defined with jet and \cPqb jet multiplicities.
Data samples enriched in background processes were used to validate predictions, as well as to constrain their uncertainties.
The larger data set and reduced systematic uncertainties such as those  associated with the lepton identification, helped to substantially improve the precision on the measured cross section with respect to previous measurements reported in Refs.~\cite{Sirunyan:2017uzs,Aaboud:2019njj}.
The measured inclusive cross section $\sigma(\ttZ)=0.95\pm0.05\stat\pm0.06\syst\unit{pb}$ is in good agreement with the standard model prediction of $0.84\pm 0.10$\unit{pb}~\cite{deFlorian:2016spz,Frixione:2015zaa,Frederix:2018nkq}.
This is the most precise measurement of the \ttZ cross section to date, and the first measurement with a precision competing with current theoretical calculations.

Absolute and normalized differential cross sections for the transverse momentum of the \PZ boson and for \cosThetaStar, the angle between the direction of the \PZ boson and the direction of the negatively charged lepton in the rest frame of the \PZ boson, are measured for the first time. The standard model predictions at next-to-leading order are found to be in good agreement with the measured differential cross sections.
The measurement is also interpreted in terms of anomalous interactions of the \PQt quark with the \PZ boson.
Confidence intervals for the anomalous vector and the axial-vector current couplings and the dipole moment interactions are presented. Constraints on the Wilson coefficients in the standard model effective field theory are also presented.

\begin{acknowledgments}
We congratulate our colleagues in the CERN accelerator departments for the excellent performance of the LHC and thank the technical and administrative staffs at CERN and at other CMS institutes for their contributions to the success of the CMS effort. In addition, we gratefully acknowledge the computing centers and personnel of the Worldwide LHC Computing Grid for delivering so effectively the computing infrastructure essential to our analyses. Finally, we acknowledge the enduring support for the construction and operation of the LHC and the CMS detector provided by the following funding agencies: BMBWF and FWF (Austria); FNRS and FWO (Belgium); CNPq, CAPES, FAPERJ, FAPERGS, and FAPESP (Brazil); MES (Bulgaria); CERN; CAS, MoST, and NSFC (China); COLCIENCIAS (Colombia); MSES and CSF (Croatia); RPF (Cyprus); SENESCYT (Ecuador); MoER, ERC IUT, PUT and ERDF (Estonia); Academy of Finland, MEC, and HIP (Finland); CEA and CNRS/IN2P3 (France); BMBF, DFG, and HGF (Germany); GSRT (Greece); NKFIA (Hungary); DAE and DST (India); IPM (Iran); SFI (Ireland); INFN (Italy); MSIP and NRF (Republic of Korea); MES (Latvia); LAS (Lithuania); MOE and UM (Malaysia); BUAP, CINVESTAV, CONACYT, LNS, SEP, and UASLP-FAI (Mexico); MOS (Montenegro); MBIE (New Zealand); PAEC (Pakistan); MSHE and NSC (Poland); FCT (Portugal); JINR (Dubna); MON, RosAtom, RAS, RFBR, and NRC KI (Russia); MESTD (Serbia); SEIDI, CPAN, PCTI, and FEDER (Spain); MOSTR (Sri Lanka); Swiss Funding Agencies (Switzerland); MST (Taipei); ThEPCenter, IPST, STAR, and NSTDA (Thailand); TUBITAK and TAEK (Turkey); NASU and SFFR (Ukraine); STFC (United Kingdom); DOE and NSF (USA).

\hyphenation{Rachada-pisek} Individuals have received support from the Marie-Curie program and the European Research Council and Horizon 2020 Grant, contract Nos.\ 675440, 752730, and 765710 (European Union); the Leventis Foundation; the A.P.\ Sloan Foundation; the Alexander von Humboldt Foundation; the Belgian Federal Science Policy Office; the Fonds pour la Formation \`a la Recherche dans l'Industrie et dans l'Agriculture (FRIA-Belgium); the Agentschap voor Innovatie door Wetenschap en Technologie (IWT-Belgium); the F.R.S.-FNRS and FWO (Belgium) under the ``Excellence of Science -- EOS" -- be.h project n.\ 30820817; the Beijing Municipal Science \& Technology Commission, No. Z181100004218003; the Ministry of Education, Youth and Sports (MEYS) of the Czech Republic; the Lend\"ulet (``Momentum") Program and the J\'anos Bolyai Research Scholarship of the Hungarian Academy of Sciences, the New National Excellence Program \'UNKP, the NKFIA research grants 123842, 123959, 124845, 124850, 125105, 128713, 128786, and 129058 (Hungary); the Council of Science and Industrial Research, India; the HOMING PLUS program of the Foundation for Polish Science, cofinanced from European Union, Regional Development Fund, the Mobility Plus program of the Ministry of Science and Higher Education, the National Science Center (Poland), contracts Harmonia 2014/14/M/ST2/00428, Opus 2014/13/B/ST2/02543, 2014/15/B/ST2/03998, and 2015/19/B/ST2/02861, Sonata-bis 2012/07/E/ST2/01406; the National Priorities Research Program by Qatar National Research Fund; the Ministry of Science and Education, grant no. 3.2989.2017 (Russia); the Programa Estatal de Fomento de la Investigaci{\'o}n Cient{\'i}fica y T{\'e}cnica de Excelencia Mar\'{\i}a de Maeztu, grant MDM-2015-0509 and the Programa Severo Ochoa del Principado de Asturias; the Thalis and Aristeia programs cofinanced by EU-ESF and the Greek NSRF; the Rachadapisek Sompot Fund for Postdoctoral Fellowship, Chulalongkorn University and the Chulalongkorn Academic into Its 2nd Century Project Advancement Project (Thailand); the Welch Foundation, contract C-1845; and the Weston Havens Foundation (USA).
\end{acknowledgments}

\bibliography{auto_generated}
\cleardoublepage \appendix\section{The CMS Collaboration \label{app:collab}}\begin{sloppypar}\hyphenpenalty=5000\widowpenalty=500\clubpenalty=5000\vskip\cmsinstskip
\textbf{Yerevan Physics Institute, Yerevan, Armenia}\\*[0pt]
A.M.~Sirunyan$^{\textrm{\dag}}$, A.~Tumasyan
\vskip\cmsinstskip
\textbf{Institut f\"{u}r Hochenergiephysik, Wien, Austria}\\*[0pt]
W.~Adam, F.~Ambrogi, T.~Bergauer, J.~Brandstetter, M.~Dragicevic, J.~Er\"{o}, A.~Escalante~Del~Valle, M.~Flechl, R.~Fr\"{u}hwirth\cmsAuthorMark{1}, M.~Jeitler\cmsAuthorMark{1}, N.~Krammer, I.~Kr\"{a}tschmer, D.~Liko, T.~Madlener, I.~Mikulec, N.~Rad, J.~Schieck\cmsAuthorMark{1}, R.~Sch\"{o}fbeck, M.~Spanring, D.~Spitzbart, W.~Waltenberger, C.-E.~Wulz\cmsAuthorMark{1}, M.~Zarucki
% ...
\vskip\cmsinstskip
\dag: Deceased\\
1:  Also at Vienna University of Technology, Vienna, Austria\\
% ...
\end{sloppypar}
\end{document}